# On the relation between magnetic field strength and gas density in the interstellar medium: A multiscale analysis

D. J. Whitworth,[1,2,3]⋆ S. Srinivasan,[1] R.E. Pudritz,[3,4] M.-M. Mac Low,[5] G. Eadie,[6,7,8] A. Palau,[1] J. D. Soler,[9] R. J. Smith,[10] K. Pattle,[11,12] H. Robinson,[4] R. Pillsworth,[4] J. Wadsley,[4] N. Brucy,[3,13] U. Lebreuilly,[14] P. Hennebelle,[14] P. Girichidis,[3] F. A. Gent,[15,16,22] J. Marin,[2] L. Sánchez Valido,[5,17] V. Camacho,[1,18] R. S. Klessen,[3,19,20,21] and E. Vázquez-Semadeni[1]

[1] *Universidad Nacional Autónoma de México, Instituto de Radioastronomía y Astrofísica, Antigua Carretera a Pátzcuaro 8701, Ex-Hda. San José de la Huerta, 58089 Morelia, Michoacán, México*
[2] *Jodrell Bank Centre for Astrophysics, Dept. of Physics and Astronomy, University of Manchester, Oxford Road, Manchester M13 9PL, UK*
[3] *Universität Heidelberg, Zentrum für Astronomie, Institut für Theoretische Astrophysik, Albert-Ueberle-Str. 2, 69120 Heidelberg, Germany*
[4] *Department of Physics and Astronomy, McMaster University, 1280 Main Street West, Hamilton, ON, L8S4K1, Canada*
[5] *Department of Astrophysics, American Museum of Natural History, 200 Central Park West, New York, NY 10024, USA*
[6] *David A. Dunlap Department of Astronomy & Astrophysics, University of Toronto, 50 St George St, Toronto, ON, M5S 3H4, Canada*
[7] *Department of Statistical Sciences, University of Toronto, 700 University Avenue, Toronto, ON M5G 1Z5, Canada*
[8] *Data Sciences Institute, University of Toronto, 700 University Avenue, Toronto, ON, M5G 1Z5, Canada*
[9] *Istituto di Astrofisica e Planetologia Spaziali (IAPS). INAF. Via Fosso del Cavaliere 100, 00133 Roma, Italy*
[10] *SUPA, School of Physics and Astronomy, University of St Andrews, North Haugh, St Andrews, KY16 9SS, UK*
[11] *Department of Physics and Astronomy, University College London, Gower Street, London WC1E 6BT, United Kingdom*
[12] *Centre for Astronomy, School of Natural Sciences, University of Galway, University Road, Galway H91 TK33, Ireland*
[13] *Centre de Recherche Astrophysique de Lyon UMR5574, ENS de Lyon, Univ. Lyon1, CNRS, Université de Lyon, 69007, Lyon, France*
[14] *Université Paris-Saclay, Université Paris Cité, CEA, CNRS, AIM, 91191, Gif-sur-Yvette, France*
[15] *HPCLab, Department of Computer Science, Aalto University, PO Box 15400, FI-00076, Espoo, Finland*
[16] *Nordita, KTH Royal Institute of Technology and Stockholm University, Hannes Alfvéns väg 12, Stockholm, SE-106, Sweden*
[17] *Department of Physics & Astronomy, Barnard College, Altschul Hall 504A, 3009 Broadway, New York, NY 10027, USA*
[18] *Center of Astronomy and Gravitation, Department of Earth Sciences, National Taiwan Normal University, 88, Sec. 4, Ting-Chou Rd., Wenshan District, Taipei 116, Taiwan R.O.C*
[19] *Universität Heidelberg, Interdisziplinäres Zentrum für Wissenschaftliches Rechnen, Im Neuenheimer Feld 225, 69120 Heidelberg, Germany*
[20] *Harvard-Smithsonian Center for Astrophysics, 60 Garden Street, Cambridge, MA 02138, USA*
[21] *Fellow at the Radcliffe Institute for Advanced Studies at Harvard University, 10 Garden Street, Cambridge, MA 02138, USA*
[22] *School of Mathematics, Statistics & Physics, Newcastle University, Newcastle upon Tyne NE1 7RU, UK*

**ABSTRACT**

The relationship between magnetic field strength $B$ and gas density $n$ in the interstellar medium is of fundamental importance. We present and compare Bayesian analyses of the $B$-$n$ relation for two comprehensive observational data sets: a Zeeman data set and 700 observations using the Davis-Chandrasekhar-Fermi (DCF) method. Using a hierarchical Bayesian analysis we present a general, multi-scale broken power-law relation, $B = B_0(n/n_0)^\alpha$, with $\alpha = \alpha_1$ for $n < n_0$ and $\alpha_2$ for $n > n_0$, and with $B_0$ the field strength at $n_0$. For the Zeeman data we find: $\alpha_1 = 0.15^{+0.06}_{-0.09}$ for diffuse gas and $\alpha_2 = 0.53^{+0.09}_{-0.07}$ for dense gas with $n_0 = 0.40^{+1.30}_{-0.30} \times 10^4$ cm$^{-3}$. For the DCF data we find: $\alpha_1 = 0.26^{+0.01}_{-0.01}$ and $\alpha_2 = 0.77^{+0.14}_{-0.15}$, with $n_0 = 14.00^{+10.00}_{-7.00} \times 10^4$ cm$^{-3}$, where the uncertainties give 68% credible intervals. We perform a similar analysis on nineteen numerical magnetohydrodynamic simulations covering a wide range of physical conditions from protostellar disks to dwarf and Milky Way-like galaxies, computed with the AREPO, FLASH, PENCIL, and RAMSES codes. The resulting exponents depend on several physical factors such as dynamo effects and their time scales, turbulence, and initial seed field strength. We find that the dwarf and Milky Way-like galaxy simulations produce results closest to the observations.

⋆ E-mail: d.whitworth@irya.unam.mx

# 1 INTRODUCTION

The role of magnetic fields in the interstellar medium (ISM) has been explored ever since their discovery more than 70 years ago (Davis 1951). Since then, magnetic fields have been found in a wide variety of astrophysical systems including galaxy clusters (e.g. Carilli & Taylor 2002), spiral galaxies (e.g. Beck 2015), the diffuse interstellar medium, giant molecular clouds (e.g. Crutcher 2012), regions of star-cluster formation (e.g. Pudritz et al. 2014; Kirk et al. 2015), and protostellar discs (e.g. Hull & Zhang 2019). It is well-recognised that magnetic fields may play a significant role in gas dynamics across many decades of physical scale which have very different properties. As an example, the question of their connection to the ISM and whether they support giant molecular clouds against collapse and affect star formation drives major observational programs (e.g. Crutcher et al. 2010, hereafter C10; Pillai et al. 2015; Planck Collaboration et al. 2016; Soler 2019) and numerical simulations (e.g. Girichidis et al. 2018; Gent et al. 2021; Ibáñez-Mejía et al. 2022; Robinson & Wadsley 2024, Whitworth et al. 2023, hereafter W23; Zhao et al. 2024) of ever growing sophistication and resolution.

Mestel (1966) first proposed a theoretical connection between magnetic field strength and density in studies of the idealised collapse of dense, magnetized, gravitationally bound spheres of gas. They found two possible power-law relationships of the form $B \propto n^\alpha$, depending on the strength of the field. In the case of weak fields with flux-freezing, the morphology of the cloud is not affected, so the field is dragged in a spherical or isotropic collapse, resulting in $|\mathbf{B}| \propto n^{2/3}$. However, stronger fields constrain collapse to only occur parallel to field lines, giving rise to a $|\mathbf{B}| \propto n^{1/2}$ scaling. Finally, fields strongly affected by ambipolar diffusion, the process where neutral particles move relative to positive ions that are tied to the magnetic field, typically yield an exponent of approximately 0.47 (Kunz & Mouschovias 2010).

The $B$-$n$ relation is far more general and plays an important role in a much larger set of astrophysical environments than just regions of isotropic gravitational collapse. While Mestel (1966) derived a relation for self-gravitating systems without turbulence, non-self gravitating, turbulent environments can also lead to a similar relation (Molina et al. 2012). As an example of the effects that turbulence can have, Passot & Vázquez-Semadeni (2003) showed that in turbulent magnetohydrodynamics (MHD) simulations without self-gravity, the relationship between magnetic pressure, $B^2$, and density depends on the type of MHD wave mode predominantly active in the flow; the fast mode produces a scaling $B^2 \propto n^2$, while the slow mode produces a scaling $B^2 \propto c_1 - c_2 n$, where $c_1$ and $c_2$ are constants. More generally, turbulence drives dynamo action that establishes field strengths in media of widely different densities, from the diffuse ISM, to the denser portion of molecular clouds. Thus Gent et al. (2021) showed that the small-scale dynamo (SSD) is driven to saturation in a turbulent multiphase diffuse medium. It is therefore important to encompass a broad range of dynamical simulations and physical processes, and to compare the $B$-$n$ relations produced with the most comprehensive observational data sets available.

C10 compiled Zeeman measurements of line-of-sight magnetic field strengths from a variety of published observational data sets. Low density gas measurements were acquired from H I and OH lines, whereas measurements at higher gas densities used OH and CN lines. Their Bayesian analysis of Zeeman measurements of 137 molecular clouds yielded a power-law relation of $B \propto n^{2/3}$ above densities of $n_0 = 300$ cm$^{-3}$, while below this density the magnetic field strength was interpreted as being independent of gas density, i.e. $B \propto n^0$.

C10 interpreted the high density power-law exponent as evidence that magnetic energy does not dominate over gravity at intermediate to high densities. This has been supported by some subsequent observations (Soler 2019) and simulations (Seifried et al. 2020; Ibáñez-Mejía et al. 2022; Whitworth et al. 2023). Other work gives different results. For example Tritsis et al. (2015) re-analysed the Zeeman data used by Crutcher, and showed that with relaxed assumptions on the errors a smaller index is found, with $B \propto n^{1/2}$ being more likely. On the other hand, Jiang et al. (2020) found a higher value for the exponent; $\alpha \sim 0.72$, and argue that due to the observational uncertainties in both $B$ and $n$ it is difficult to accurately determine the exponent and density at which the turn over occurs.

An important recent development is that observational data sets have been greatly expanded with submillimeter polarization measurements in denser regions using instruments like the Atacama Large Millimeter-submillimeter Array or the Submillimeter Array (SMA; Pattle et al. 2023). Observations of polarized thermal dust emission and density, along with the velocity dispersion $\sigma_\mathrm{v}$, provide estimates of the plane-of-sky magnetic field strength by the Davis-Chandrasekhar-Fermi (DCF) method (Davis 1951; Chandrasekhar & Fermi 1953). These estimates are performed using a combination of the scatter of polarization directions and the strength of turbulence implied by the velocity dispersion.

The main assumptions of the DCF method are: 1) that there is an underlying uniform magnetic field perturbed by the turbulent gas motions, 2) that the dispersion in polarization position angles, $\sigma_\mathrm{PA}$, is produced by the propagation of incompressible MHD waves (i.e., Alfvén waves), and 3) that the gas kinetic energy (which is assumed to be fully turbulent) is completely transferred to magnetic energy fluctuations. The DCF method allows the determination of the plane-of-sky component of the magnetic field strength

$$B_\mathrm{pos} = f(4\pi\rho)^{1/2} \frac{\sigma_\mathrm{v}}{\sigma_\mathrm{PA}}, \quad (1)$$

where the gas mass density is $\rho$ and $f$ is a correction factor further discussed below. An initial analysis of DCF data fitting a single power-law to the $B$-$n$ relation using a least-squares fit found $\alpha \sim 0.57$ (Liu et al. 2022b).

One source of uncertainty in the DCF method is that whereas it has been generally assumed that $\sigma_\mathrm{PA}$ is produced by the propagation of incompressible MHD waves, compressible modes of turbulence could *also* make a significant contribution. Skalidis & Tassis (2021, henceforth ST) and Skalidis et al. (2021) develop a new method to take the compressible modes into account. This method is new and has yet to be rigorously tested observationally and numerically but the early results are promising.

The large body of numerical work now available is transforming our understanding of the $B$-$n$ relation. On the scale of dwarf galaxies W23 reported a relationship of $|\mathbf{B}| \propto n^{1/2}$ in high resolution simulations. Other theoretical work also shows a similar exponent (Hennebelle et al. 2008; Banerjee et al. 2009; Girma & Teyssier 2023). However, in different simulations the *B-n* relation can vary depending on numerous conditions such as initial seed field strength (Robinson & Wadsley 2024), dynamo effects (Gent et al. 2024; Korpi-Lagg et al. 2024), embedded ISM dynamics (Ibáñez-Mejía et al. 2022), and time dependencies (Robinson & Wadsley 2024; Konstantinou et al. 2024). These can all affect whether a break in the relation appears at some density and the value of the exponents.

Other recent theoretical magneto-gravo-turbulent simulations (Mocz et al. 2017; Auddy et al. 2022) find that the power-law exponent for diffuse gas in isolated molecular clouds increases as the Alfvénic Mach number increases. Models of individual, isolated, collapsing cores by Cao & Li (2023) show how their *B-n* relationship evolves. At the end of the simulations the cores match the C10 result, suggesting that it is driven by the gas being magnetically super-critical with super Alfvénic turbulence (Li et al. 2013; Myers & Basu 2021).

The goal of this paper is to address two basic questions. What is the observationally derived general magnetic field - density relation across a wide range of ISM densities? What are the likely physical processes driving it, determined through simulations? We present a new multi-scale, hierarchical Bayesian analysis of the relationship for two extended data sets; Zeeman and DCF observations. We also perform a non-hierarchical Bayesian analysis of a comprehensive suite of simulations performed by our team of a wide variety of regions in the ISM—from protostellar discs to entire galaxies—using a variety of MHD codes. This is done in order to ascertain what physical processes account for the observed relation, including such issues as what affects the value of the break point density $n_0$ and related field strength $B_0$.

In general, these simulations show that up to densities of $10^5$ cm$^{-3}$, ideal MHD applies: the field is tied to the flow of the gas as the effects of ambipolar diffusion are subdominant (e.g. Mac Low & Klessen 2004). For galaxy scale simulations, ideal MHD is sufficient as they do not resolve down to scales where non-ideal effects become important. Simulations that model protostellar discs (Lebreuilly et al. 2021) and individual star formation (Moscadelli et al. 2022) include non-ideal MHD physics at densities where ambipolar diffusion becomes important.

Section 2 briefly describes the methods used in our analysis; Section 3 shows the results from recent observations; Section 4 shows the results from simulations using different initial conditions and codes; Section 5 discusses the results of the observational parameters and the numerical simulations looking at physical processes that drive the relationship; and Section 6 summarises our findings.

## 2 METHODS

In this section we briefly describe our methods, leaving many of the details to a series of Appendices. We first introduce the functional form we assume for the *B-n* relation (Sect. 2.1) and describe our Bayesian analysis in Section 2.2; Appendix A provides a more detailed description of the analysis and error treatment. We then outline the observational and numerical data sets used in this study, along with their uncertainties, in Sections 2.3 and 2.4.

### 2.1 Functional Form

In order to compare and contrast with the original C10 relation, we assume a two-part broken power-law *B-n* relation for our model, similar to theirs;

$$|\mathbf{B}| = B_0 \begin{cases} (n/n_0)^{\alpha_1} & \text{if } n \leq n_0, \\ (n/n_0)^{\alpha_2} & \text{if } n > n_0. \end{cases} \quad (2)$$

where $\alpha_1$ and $\alpha_2$ are the exponents of the two power-laws, $n_0$ is the number density of the break in the power-law, and $B_0$ is the magnetic field strength at the break. Our model differs from C10 in that $\alpha_1$ is not set to zero. We do not consider a maximum on *B* but use the reported values in their respective publications. We discard the flexibility function free parameter that was used in C10 which denotes the trust in the probability density function (PDF) used in the prior. As described in the following section, we also consider a more detailed hierarchical Bayesian approach.

We considered using a three-part power law, since including DCF observations yields larger data sets that enable stronger statistical analyses and cover a larger density range, up to $n \simeq 10^9$ cm$^{-3}$. Pattle et al. (2023) note that at densities of $n > 10^7$ cm$^{-3}$ the field strength appears to become constant, as we show in Appendix A5. However, above this density, the DCF method becomes unreliable for various reasons. For example, in this regime it is expected that dust grains will align more to the radiation field than to the magnetic field (Lazarian 2020). Therefore, we do not use the three-part power law.

### 2.2 Bayesian Analysis

In this work, we perform Bayesian inference on three different data sets: two observational, and one based on our large set of simulations. For the observational data, we use a hierarchical Bayesian model[1] and for the simulations, we use a non-hierarchical Bayesian model.

The details of the hierarchical Bayesian model for the two observational data sets are given in Appendix A2. The hierarchical Bayesian model allows us to:

(i) model the measurements of $n$ and $B$ as draws from a distribution centred on their true but unknown values,
(ii) carefully and coherently account for the measurement uncertainties in $n$ and $B$,
(iii) comprehensively study the *covariance* in measurement uncertainties inherent to the DCF data set, and
(iv) introduce physically motivated but not too-constrained priors on the model parameters.

In Section 2.3, we give an overview of how the uncertainties are incorporated into the hierarchical Bayesian model. In all forms of Bayesian analysis, the choice of the prior is important and can heavily influence the results. Appendix A2 provides a more detailed description of the likelihood, priors, and hyperpriors.

One point that has been raised in the literature (Tritsis et al. 2015; Jiang et al. 2020) is the value C10 chose for the error of $n_0$. In the original survey data no errors were reported and so a relative uncertainty of a factor of two was chosen. In Appendix A3 we test three different values for the relative uncertainty, 2, 5, and 9. These are based on the original choice, the best result from Jiang et al. (2020) of nine, and an intermediate value. We find there is some variation in the results when we change the uncertainty, but it is not statistically significant in our analyses, so we choose a factor of 5.

---

[1] Initially, we tried a non-hierarchical model for the observational data. However, we found that excluding the information about the covariance structure in $n$ and $B$ measurements led to bimodal posterior distributions (see Appendix A1.1).

The third data set we analyse is the ensemble of all of our numerical simulations. Here we employ a broad uniform PDF of priors for all parameters due to the large range of densities covered in the simulations, $10^{-5} < n_0 < 10^{13}$ cm$^{-3}$ and the resulting spread of data in each. Due to the amount of data in each simulation—over $10^6$ data points in each model—we use a non-hierarchical Bayesian model for computational efficiency. We perform inference on five random samples of $10^5$ data points from each simulation and combine the results to compute credible intervals[2] for the parameters. A detailed discussion of the non-hierarchical model used can be found in Appendix A6.

### 2.3 DCF errors

When $B_{\text{pos}}$ is inferred from the DCF method, the main uncertainties come from the following sources:

(i) Density: The estimate of the density in Eq. (1) is usually obtained by combining a mass estimate derived from thermal dust emission with the assumption of a specific geometry. To estimate the mass, a specific dust temperature and a dust opacity at the observing frequency are usually assumed (e.g., Könyves et al. 2015; Elia et al. 2017). These two quantities are highly uncertain, with the dust opacity varying by a factor of 2–4 (e.g., Ossenkopf & Henning 1994; Li et al. 2024).

(ii) Velocity dispersion: The velocity dispersion is measured from a molecular gas tracer. It is typically assumed that the entire velocity dispersion, which is usually dominated by non-thermal motions, is due to turbulence only. Therefore, the uncertainty comes from the specific properties of the tracer used, and from the fact that a non-negligible part of the velocity dispersion might be due to bulk, non-turbulent, motions. If this is not taken into account, the magnetic field can be overestimated by a factor of 2, as shown in Palau et al. (2021).

(iii) Polarization position angle dispersion: When a number of polarization segments have been detected in a specific region, $\sigma_{\text{PA}}$ can be characterized with different measures, such as the standard deviation of the position angles, the width of a Gaussian fit to the position angle histograms, or more sophisticated values based on the structure function as proposed by Hildebrand et al. (2009) or Houde et al. (2009). Even if the structure functions account for the mean-field orientation, they can be strongly affected by the sampling in the polarization data (e.g., Soler et al. 2016, see below). These different approaches to estimate $\sigma_{\text{PA}}$ typically differ by a factor of two (e.g., Palau et al. 2021).

(iv) $f$ factor: The correction factor $f$ has been proposed to correct for several biases, such as projection effects along the line-of-sight, the averaging of the magnetic field within the beam, the use of only one of the three velocity components, and the lack of exact equipartition between the magnetic energy of the perturbed magnetic field and the kinetic energy. This factor has been estimated from numerical simulations to usually be 0.5, although it can reach lower values of around 0.2 (e.g., Heitsch et al. 2001; Ostriker et al. 2001; Liu et al. 2022a).

(v) Sparse sampling effects: Due to sensitivity limitations, only a fraction of the total area emitting polarized radiation is detected, restricting the reliability of the measurement. An extreme case is when very few polarization segments are detected, suggesting a relatively uniform and strong field, while deeper observations reveal an underlying more turbulent field. The relative uncertainty due to sparse sampling can reach up to 60% (Palau et al. 2021). This uncertainty is explained in detail in Liu et al. (2019).

(vi) Cloud extent along the line of sight (LOS): Usually, there is limited information on the target's size in the LOS direction. This uncertainty can lead to considerable error in the density estimations. For example, Angarita et al. (2024) shows that clouds could be up to ten times deeper than predicted through plane-of-sky observations.

In addition to the aforementioned uncertainties, the choice of the DCF or the ST method for interpretation of the polarization data can introduce additional biases. The main difference between the DCF and ST methods is the dependence on $\sigma_{\text{PA}}$ (apart from the $\sqrt{2}$ factor): in the DCF method, $B_{\text{pos}} \propto 1/\sigma_{\text{PA}}$, while in the ST method, $B_{\text{pos}} \propto 1/\sqrt{\sigma_{\text{PA}}}$. Since in most cases $\sigma_{\text{PA}} < 1$ rad, the ST method results in $B_{\text{pos}}$ systematically smaller by a factor of 2–4 compared to the values obtained with the DCF method.

Taking into account the combination of all the potential uncertainties associated with the $B_{\text{pos}}$ measurements, the global uncertainties could reach an order of magnitude. Further details on the uncertainties associated with magnetic field measurements can be found in the recent reviews by Liu et al. (2021), Liu et al. (2022a, see their Sec. 2.2), and Pattle et al. (2023).

### 2.4 Observational data sets

The Zeeman splitting data of C10 (their Table 1), used for their analysis, provides estimates of the line-of-sight component of the magnetic field. For the DCF data in this work, we used the compilation of Pattle et al. (2023).

The data is split into two different sets in order to provide a statistical analysis consisting of the original Zeeman data along with a new data set based on the DCF measurements from Pattle et al. (2023). These are:

**1** Zeeman: The Zeeman data from C10.
**2** DCF: The DCF data from Pattle et al. (2023) with only symmetric errors in both $n$ and B.

In addition to the two main data sets we tested different subsets of the data. First we tested whether unrealistically small error estimates were biasing the results. To do this we set any error that is smaller than 15% to 15%, and the same for 50%. We saw no statistically significant change in the results, and so do not include this analysis. We also tested whether an outlier in the DCF data set was biasing our results (see Appendix A5). We test the scaling ratio of Pattle et al. (2023) between DCF data and Zeeman data in Appendix A4 but due to inherent uncertainties in this we do not consider this a main result. Finally, we also looked at the full data set from Pattle et al. (2023) with the known errors plus errors computed using a kernel density estimate (KDE) for the data that was lacking reported errors (see Appendix A3.3 for a discussion on the errors). We report these results in Appendix A5, as we consider them less reliable due to the inherent uncertainty in the KDE error estimates.

---

[2] We compute two types of 68% credible intervals: equal-tailed intervals which reject 16% of the total probability on either side of the median, and highest posterior density (HPD) intervals that select the narrowest interval containing 68% of the total probability. For unimodal distributions, HPD intervals will contain the mode of the distribution.

### 2.5 Numerical simulations: methods and data sets

We perform the non-hierarchical Bayesian analysis on the ensemble of nineteen numerical simulations that cover a wide range of initial conditions, densities, magnetic field strengths, and scales. The systems covered in our simulations range from full Milky Way-like galaxies to protoplanetary discs.

The simulations themselves were executed with the codes AREPO (Springel 2010), FLASH (Fryxell et al. 2000), RAMSES (Teyssier 2002), and PENCIL (Brandenburg & Dobler 2002; Brandenburg et al. 2020). These codes use different types of meshes and refinement criteria. AREPO solves the MHD equations on a quasi-Lagrangian Voronoi moving mesh that provides high spatial resolution in high-density regions. FLASH and RAMSES are adaptive mesh refinement (AMR) codes that use hierarchical nested Cartesian grids to achieve high resolution in regions of interest. PENCIL is a high-order, uniform grid code using sixth-order space and fourth-order time derivatives. We refer the reader to the cited method papers for a more detailed description of each code.

For each simulation presented here, we refer to the original publication for a detailed discussion of their numerical setup. The high-resolution simulations of dwarf galaxies presented here are novel and have been run specifically for this work, so we summarize the key differences to the original simulations in Appendix B.

For each simulation, we plot the distribution of $|\mathbf{B}|$ against the number density $n$ in each computational cell, the derived non-heirachical bayesian parameters and the original relation of C10. The volume-weighted average fit to the data is calculated by binning the data into 500 equally sized log-spaced density bins and then calculating the volume-weighted average of each bin.

We volume weight the magnetic field. In the uniform grid code this is most comparable to the Eulerian nature of the code. For the AREPO and AMR simulations, we volume weight to avoid oversampling in the densest regions. The meshes are non-uniform and so a simple grid average would be biased towards the densest regions. By volume weighting, we take an Eulerian point of view, meaning that we get the typical view at a randomly chosen point in space.

## 3 OBSERVATIONAL RESULTS

Figure 1 displays the joint distribution of the posterior samples from our hierarchical Bayesian analysis of the observational data. Both the joint and marginal parameter distributions for the Zeeman and DCF data are shown. The distributions show some strong correlations between parameters, which is expected. For example, the relationship between $n_0$ and $B_0$ and the relationship between $\alpha_1$ and $B_0$ are particularly strong.

Using the samples from the posterior distributions in Figure 1, the inferred relationships between $B$ and $n$ are illustrated in Figure 2. It shows two generalised forms of the relationship, one based on the Zeeman data from C10 and the other based on the DCF data from Pattle et al. (2023). The first of these compares the data with the inferred maximum a posteriori (MAP) model constructed using posterior samples for the Zeeman data. The second compares the DCF data set both with the MAP model and with the range of posterior predictions for this data set, demonstrating that the models can reproduce the range of observed values.

Before going further, we want to stress a key difference between the bottom two panels in Figure 2, because this is important for interpretation. If one wishes to read the mode of the posterior as the most likely values of $(\alpha_1, \alpha_2, n_0, B_0)$, then one should look at the left-hand panel. The red line corresponds to the unique set of parameter values $(\alpha_1, \alpha_2, n_0, B_0)$ at the MAP. This is a *point estimate* of the posterior mode derived from our samples from the posterior distribution; this line does not correspond to any particular draw from the posterior.

A more Bayesian perspective is presented in the right-hand panel, which shows the pointwise predicted relationship of magnetic field to number density, as predicted by the posterior distribution. The blue region is the 68% Bayesian credible region, which encompasses our uncertainty in the prediction given the data, our priors, and the assumed model. The pointwise predicted relationship is calculated as follows:

**1** At each value of $n$, the value of $B$ is calculated for every sample $(\alpha_1, \alpha_2, n_0, B_0)$ from the posterior distribution. Because we have $5\,000 \times 4 = 20\,000$ samples from the posterior, this results in $20\,000$ values of $B$.

**2** For each $n$, the median of the $B$ values is calculated and plotted in purple, resulting in the relationship shown.

**3** For each $n$, the $20\,000$ values of $B$ from step **1** are used to compute $16^{\text{th}}$ and $84^{\text{th}}$ percentiles to construct the 68% equal-tailed interval, shown in blue.

Table 1 summarises the estimated median and mode along with the corresponding 68% credible intervals for each parameter, calculated from the posterior samples. The exponent in the low-density regime, $\alpha_1$, is tightly constrained and positive in all cases. The values derived for the two data sets are consistent within their 68% confidence intervals. The corresponding field strength normalisation $B_0$ is not consistent between datasets. This is not surprising because there is a strong correlation or degeneracy between $B_0$ and $n_0$ (Fig. 1). We discuss this further in Section 5.2.

In our analysis, we note that we have excluded one data point in the DCF data that lies far beyond $n = 10^7$ cm$^{-3}$. We conducted our analysis with and without this data point and found that it was highly influential (see Appendix A5).

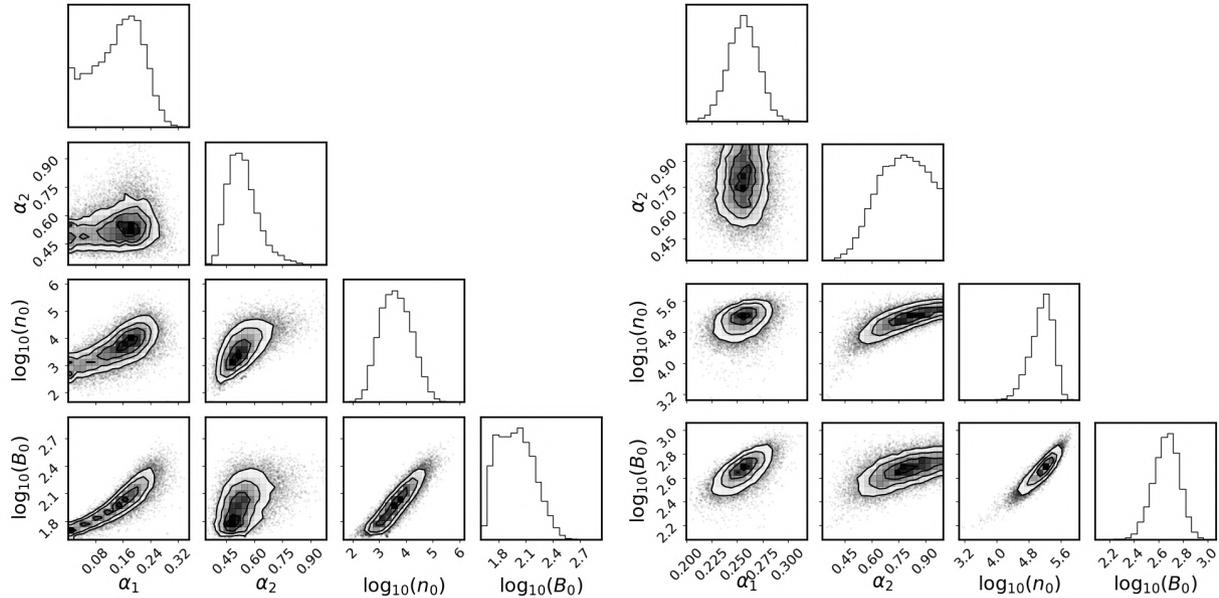

**Figure 1.** The joint and marginal parameter distributions produced by the hierarchical Bayesian model for the Zeeman (*left*) and DCF (*right*) data sets. Almost all joint distributions show a significant correlation between the parameters.

This influential data point also leads us to explore the influence of including many high-density values. In Appendix A5, we perform an analysis on the complete DCF data set and find that the results become unphysical with the high-density values included.

Figure 3 shows the median and MAP parameter estimates of the posterior samples and the corresponding 68% credible intervals for the two observational data sets as well as each of the nineteen numerical simulations examined in this work. The medians are estimated from the combined posterior samples from each data set with errors propagated through. The data without visible error bars around the median and MAP have narrow, but non-zero, credible errors.

The Zeeman and DCF data agree in having $\alpha_1 > 0$ in the diffuse gas, $\alpha_2 > 0.5$ in the dense gas, a break density $n_0 > 10^3$ cm$^{-3}$, and $B_0 > 50\,\mu$G. These are useful points of comparison to the models, despite the remaining substantial variation in the values for each estimate of the $B$–$n$ relationship.

| Data set | $\alpha_1$ | $\alpha_2$ | $n_0$ ($10^4$ cm$^{-3}$) | $B_0$ ($\mu$G) |
|---|---|---|---|---|
| **Zeeman** | | | | |
| Median | $0.15^{+0.06}_{-0.09}$ | $0.53^{+0.09}_{-0.07}$ | $0.40^{+1.30}_{-0.30}$ | $95.00^{+60.00}_{-36.00}$ |
| MAP | $0.18^{+0.05}_{-0.10}$ | $0.50^{+0.10}_{-0.10}$ | $0.26^{+0.53}_{-0.25}$ | $61.00^{+63.00}_{-14.00}$ |
| **DCF** | | | | |
| Median | $0.26^{+0.01}_{-0.01}$ | $0.77^{+0.14}_{-0.15}$ | $14.00^{+10.00}_{-7.00}$ | $460.00^{+120.00}_{-100.00}$ |
| MAP | $0.26^{+0.01}_{-0.02}$ | $0.78^{+0.15}_{-0.14}$ | $12.00^{+8.00}_{-8.00}$ | $440.00^{+130.00}_{-93.00}$ |

**Table 1.** Median parameter estimates for the observational $B$-$n$ relation with 68% equal-tailed intervals calculated from the posterior samples, along with the maximum a posteriori (MAP) value with the corresponding 68% highest posterior density (HPD) intervals. Both data sets have a tightly constrained, positive median $\alpha_1$ and higher $n_0$ than that reported in C10 in both the median and MAP results.

## 4 NUMERICAL SIMULATION RESULTS

In this section, we use a non-hierarchical Bayesian approach to deduce the $B$-$n$ relations from a heterogeneous set of numerical simulations using the different codes, scales, and approximations described in Section 2.5. Table 2 lists the codes used, systems simulated, highest resolution, density range covered, and whether self-gravity is included. Because the numerical simulations span a much larger range of $n$ than the observational data, as well as being much larger data sets, we use the non-hierarchical Bayesian approach with uniform priors described in Appendix A6. Figure 3 shows the parameters derived for each simulation, and Appendix A8 presents these same data in tabular form. We provide a detailed description of each simulation in the following subsections.

Examining the results, we find large variations in all free parameters across the simulations, for example, by a factor of over $10^4$ in $n_0$. Although no simulation completely matches the observational result, we find that the galactic scale simulations are most similar, with MHD_SAT_HR being the closest match for the exponents. However, only one simulation, MHD_10, has values of both $n_0$ and $B_0$ that lie within $1\sigma$ of the C10 relationship. The exponent of the diffuse gas for each simulation $\alpha_1$ varies from $-0.27$ in IMHD to $0.84$ in MHD_10_LR. Most simulations that trace the dense gas well show $\alpha_2 > 0.5$. The exceptions are the simulations run for shorter time scales like B3, B6, KPC_LOW, KPC_HIGH, and the dynamo models of Gent et al. (2021) and Gent et al. (2024), which do not reach sufficiently high densities. We discuss the details of these different simulations in the following subsections.

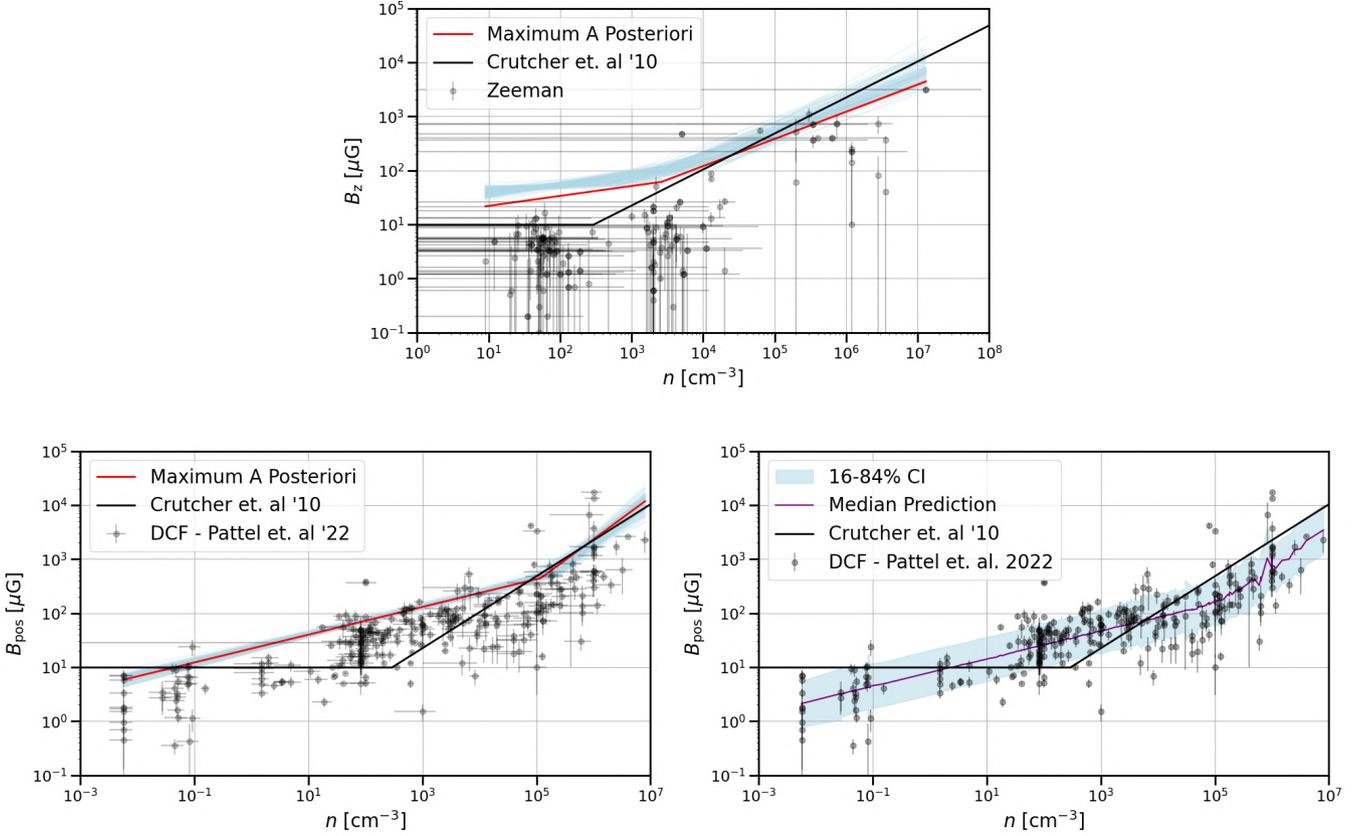

**Figure 2.** The inferred relationship of magnetic field strength as a function of number density, based on a Bayesian analysis of the observations. The *solid black line* in all plots is the relationship proposed by C10. *Top Panel:* The inferred MAP relationship between $B_z$ and $n$ based on our hierarchical Bayesian analysis applied to the Zeeman data (*red line*) from C10, along with the relationship predicted by 100 random samples from the posterior (blue lines). We include only a sample of the error bars for $n$ for readability

. *Bottom Panels:* The inferred $B_\mathrm{pos}$–$n$ relationship, based on our hierarchical Bayesian analysis applied to the DCF data set from Pattle et al. (2023). The bottom left panel shows the MAP relationship from the posterior (*red line*) and 100 random samples from the posterior (*blue lines*). The bottom right panel shows the posterior predictive plot: the median pointwise predicted field for each density value in the DCF data (*purple line*) and the 68% credible interval around this median (*blue region*), which reproduces the range of observations (*circles with error bars*). A description of the method for producing the pointwise predictive plot is given in the text.

### 4.1 Core collapse simulations

We begin by presenting results from simulations of molecular cloud cores collapsing to form protoplanetary discs and envelopes, computed with the AMR MHD code RAMSES including non-ideal effects (Lebreuilly et al. 2021). The resolution of these simulations goes from 0.0120 pc (2460 au) in the most diffuse gas to $5.84 \times 10^{-6}$ pc (1.20 au) in the densest gas. A uniform seed field of $\sim 9.40 \times 10^{-5}$ G is initialised in the core. For more detail we refer to the original publication.

Figure 4 presents data from simulations in either the ideal MHD limit or including non-ideal ambipolar diffusion. These simulations include higher density gas characteristic of protostellar discs than is possible to measure in the observed systems. There appears to be a well-defined $B$-$n$ relation with a best-fitting power-law index of $\alpha_2 \sim 0.56$. The diffuse gas index $\alpha_1$ has a negative value as the simulations only include small amounts of lower-density gas in outflow cavities unrepresentative of the ISM. We see that the fits to the two simulations are similar. This suggests that at the densities modelled, ambipolar diffusion has not strongly decoupled the ions and the field from the bulk of the gas. The drop in the power law towards a plateau in the NI model at the very highest densities may represent the first indication of substantial decoupling at high densities ($n > 10^9$ cm$^{-3}$). This decoupling behaviour is well documented (Masson et al. 2016; Vaytet et al. 2018; Hennebelle et al. 2020; Commerçon et al. 2022).

### 4.2 Stratified box simulations

We next present results from stratified, horizontally periodic domains with and without galactic shear. Each simulation represents a region of the ISM and includes supernova (SN) driving and a disc gravitational potential. All simulations except those of Gent et al. (2021) and Gent et al. (2024) include self-gravity. For a more detailed discussion of the initial conditions of each simulation, we point the reader to the original publications.

Figure 5 shows the results from two $(0.5 \text{ kpc})^3$ stratified box simulations taken from the SILCC simulations of Girichidis et al. (2018) with the FLASH AMR code. We show simulations B3 and B6, run with initial seed fields of 3 $\mu$G and 6 $\mu$G, respectively. Both simulations have a base spatial resolution of 4 pc, going down to 1 pc in the densest gas, and have been run for 60 Myr.

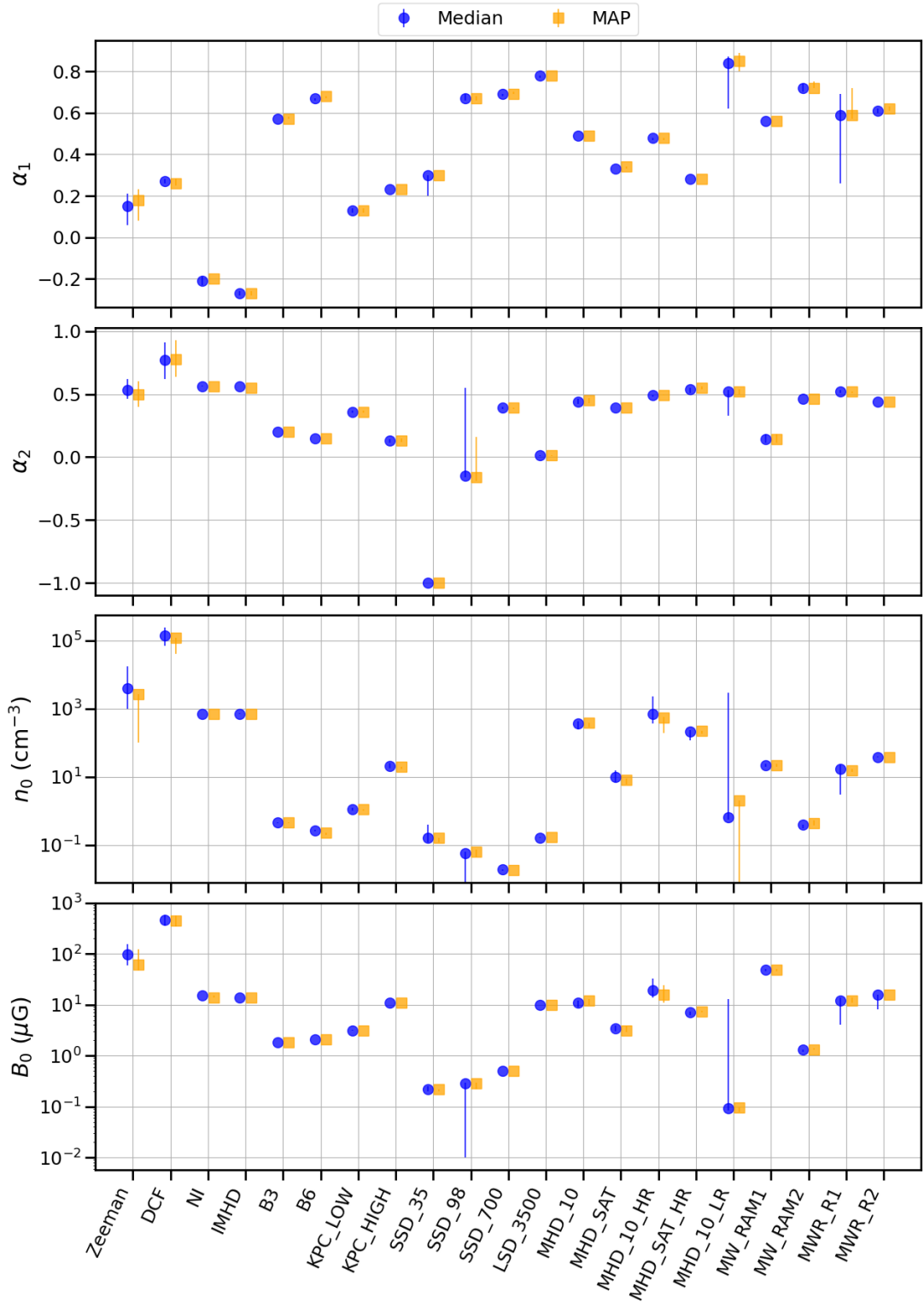

**Figure 3.** MAP (*orange squares*) and median (*blue circles*) parameter estimates for the two observational data sets and all nineteen numerical simulations. Error bars give the 68% credible intervals where they are larger than the symbol size. A tabulation of these data is available in Appendix A8.

| Simulation | Code | System | Min $\Delta x$ (pc) | Density range (cm$^{-3}$) | Self-gravity | Reference |
|---|---|---|---|---|---|---|
| NI | RAMSES | Core Collapse | $5.84 \times 10^{-6}$ | $10^1$–$10^{13}$ | Yes | Lebreuilly et al. (2021) |
| IMHD | RAMSES | Core Collapse | $5.84 \times 10^{-6}$ | $10^1$–$10^{13}$ | Yes | Lebreuilly et al. (2021) |
| B3 | FLASH | Stratified Box | 1 | $10^{-6}$–$10^5$ | Yes | Girichidis et al. (2018) |
| B6 | FLASH | Stratified Box | 1 | $10^{-6}$–$10^5$ | Yes | Girichidis et al. (2018) |
| KPC_LOW | RAMSES | Stratified Box | 4 | $10^{-4}$–$10^4$ | Yes | Brucy et al. (2023) |
| KPC_HIGH | RAMSES | Stratified Box | 0.12 | $10^{-4}$–$10^4$ | Yes | Colman et al. (2022) |
| SSD_35 | PENCIL | Stratified Box + Shear | 1 | $10^{-4}$–$10^2$ | No | Gent et al. (2021) |
| SSD_98 | PENCIL | Stratified Box + Shear | 1 | $10^{-4}$–$10^2$ | No | Gent et al. (2021) |
| SSD_700 | PENCIL | Stratified Box + Shear | 1 | $10^{-4}$–$10^2$ | No | Gent et al. (2021) |
| LSD_3500 | PENCIL | Stratified Box + Shear | 4 | $10^{-4}$–$10^0$ | No | Gent et al. (2024) |
| MHD_10 | AREPO | Dwarf Galaxy | 0.1 | $10^{-4}$–$10^5$ | Yes | W23 |
| MHD_SAT | AREPO | Dwarf Galaxy | 0.1 | $10^{-4}$–$10^5$ | Yes | W23 |
| MHD_10_HR | AREPO | Dwarf Galaxy | 0.01 | $10^{-5}$–$10^8$ | Yes | New simulation |
| MHD_SAT_HR | AREPO | Dwarf Galaxy | 0.01 | $10^{-5}$–$10^8$ | Yes | New simulation |
| MHD_10_LR | AREPO | Dwarf Galaxy | 0.1 | $10^{-4}$–$10^4$ | Yes | W23 |
| MW_RAM_1 | RAMSES | Milky Way-like Galaxy | 9.15 | $10^{-4}$–$10^3$ | Yes | Robinson & Wadsley (2024) |
| MW_RAM_2 | RAMSES | Milky Way-like Galaxy | 9.15 | $10^{-4}$–$10^3$ | Yes | Robinson & Wadsley (2024) |
| MWR_R1 | RAMSES | MW-like Galaxy Zoom-in | 0.286 | $10^{-4}$–$10^6$ | Yes | Zhao et al. (2024) |
| MWR_R2 | RAMSES | MW-like Galaxy Zoom-in | 0.286 | $10^{-4}$–$10^6$ | Yes | Zhao et al. (2024) |

**Table 2.** Simulations used. In the grid codes the maximum resolution is the size of the smallest grid cell in the dense gas. For the AREPO simulations, due to the adaptive nature of the mesh, we report the cell radius at which sink particles are formed. The cell masses in the AREPO simulations are $500\,\mathrm{M}_\odot$ for model MHD_10_LR, $50\,\mathrm{M}_\odot$ for the other models from W23, and $1\,\mathrm{M}_\odot$ in the new high resolution models.

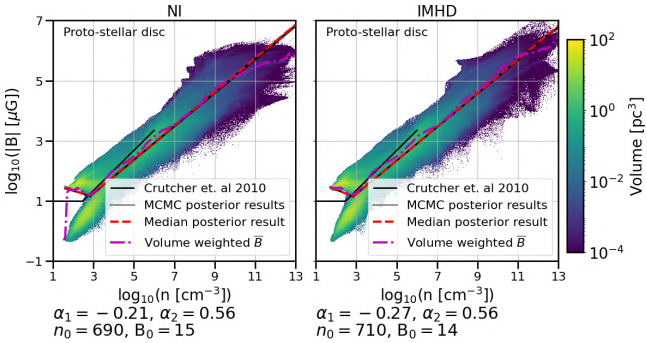

**Figure 4.** Core collapse to disc simulations: the volume-weighted absolute magnetic field $|\mathbf{B}|$ and number density $n$ distribution from *(left)* the non-ideal and *(right)* the ideal MHD simulation of Lebreuilly et al. (2021). The *red dashed line* is the relation computed using the median parameter values inferred from the five random sample sets. The *grey lines* show the best-fitting relations from the five independent random samples from the data set, each consisting of 100,000 points as described in the text. These show little scatter and lie underneath the red line. The *black line* represents the C10 relation. The *purple dashed dotted line* shows the volume-weighted average B-field strength binned by density. We can see that the relation inferred from our Bayesian analysis is similar to the volume-weighted mean B-field curve, and they are both close to $\alpha_2 \simeq 0.56$. We note the values for each parameter under the related plot.

Comparing the two simulations, we find that $n_0 < 0.5\,\mathrm{cm}^{-3}$ and that they have a diffuse gas exponent of $\alpha_1 = 0.57$–$0.67$. In the dense gas the exponent is reduced; $\alpha_2 = 0.20$ in B3 and $\alpha_2 = 0.15$ in B6, in contrast to the C10 relationship.

Figure 6 shows the results from Colman et al. (2022) and Brucy et al. (2023) run using the AMR code RAMSES to compute MHD simulations in a $(1\,\mathrm{kpc})^3$ domain. The two simulations include one with a uniform cell resolution of 4 pc and another with an AMR grid ranging from 4 pc in the low-density gas to 0.12 pc in the high-density. The lower and uniform resolution simulation KPC_LOW was run for 58 Myr while the AMR simulation KPC_HIGH was run for an additional 30 Myr at uniform resolution before being allowed to refine to higher resolution and then run for an additional 20 Myr. The initial magnetic field in both simulations was set to 7 $\mu$G.

The exponents in both the diffuse and dense gas of KPC_LOW are small, $\alpha_1 = 0.13$ and $\alpha_2 = 0.36$. In KPC_HIGH the diffuse gas exponent is larger, $\alpha_1 = 0.23$, whilst the dense gas exponent is rather lower, $\alpha_2 = 0.13$. The break occurs between $n_0 \sim 1\,\mathrm{cm}^{-3}$ and $n_0 \sim 21\,\mathrm{cm}^{-3}$, one to two orders of magnitude below the break expected from C10.

Figure 7 shows results from the PENCIL simulations of Gent et al. (2021) and Gent et al. (2024) taken at various times to study the effect of the SSD and the large-scale dynamo (LSD), as described in the review by Korpi-Lagg et al. (2024). These simulations include SN feedback, heating and cooling, and, in the case of the LSD simulation, vertical stratification and galactic shear. The SSD simulation has a resolution of 1 pc, while the LSD simulation has a resolution of 4 pc. We plot the $B$–$n$ relationship at three times for the SSD simulation: just as the field begins to grow at 35 Myr; when the SSD begins to saturate at 98 Myr; and after the SSD has fully saturated, but before the LSD has had time to grow. We plot the LSD simulation at 3500 Myr, when the LSD is approaching saturation.

Over time, starting in the diffuse medium, a positive $\alpha_1$ begins to appear. By 98 Myr (SSD_98) $\alpha_1 \sim 0.67$ with $\alpha_2$ still slightly negative, likely a result of the initial conditions. At $\sim 700$ Myr (SSD_700) $\alpha_1 \sim 0.69$, almost unchanged, and the break density remains similar as well at $n_0 \sim 0.019\,\mathrm{cm}^{-3}$. The LSD eventually saturates in LSD_3500, where $\alpha_1 = 0.78$ and $\alpha_2 = 0.012$, with a break at a slightly higher density of $n_0 \sim 0.16\,\text{cm}^{-3}$. The value of $B_0$ has increased by a factor of $\sim 40$, to the value of $10\,\mu$G seen in C10.

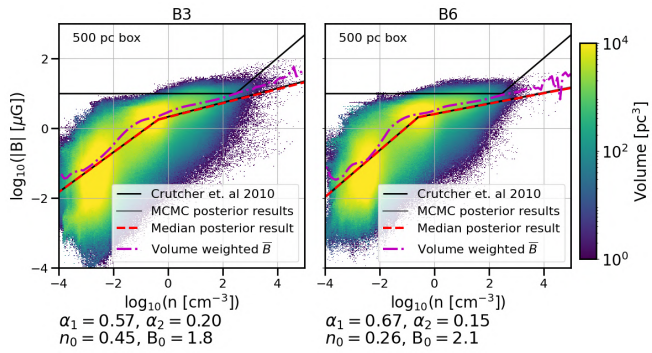

**Figure 5.** Stratified $(0.5\,\text{kpc})^3$ cube simulations without shear from Girichidis et al. (2018) with initial field strength of *(left)* $3\,\mu$G and *(right)* $6\,\mu$G. The same notation is used as in Figure 4. All of the data in these simulations lie below the Crutcher relationship with a low power-law exponent in the denser gas, a transition at $n \sim 0.3\,\text{cm}^{-3}$, and a larger power-law exponent at low densities.

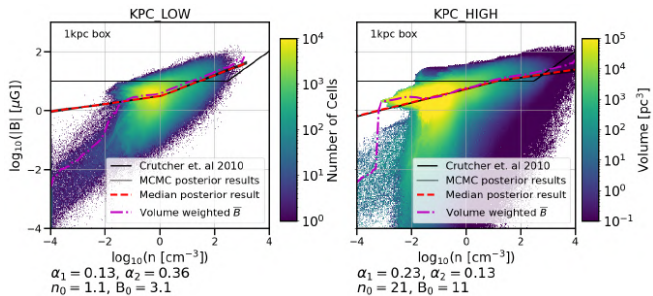

**Figure 6.** Stratified $(1\,\text{kpc})^3$ cube simulations without shear. The same notation is used as in Figure 4. *(Left)* the uniform 4 pc resolution simulation of Brucy et al. (2023); *(right)* the AMR simulation with highest resolution of 0.12 pc of Colman et al. (2022). These simulations do not agree well with the C10 relationship, with only the lower resolution simulation having a larger exponent in the denser gas. There is more scatter in the results in the diffuse gas. The high-resolution simulation appears to have almost a single slope.

### 4.3 Isolated galactic simulations

We finally turn to simulations of entire isolated galaxies, run with AREPO for dwarf galaxies and with RAMSES for galaxies with mass approaching the Milky Way. These simulations all include self-gravity and live dark matter halos.

#### 4.3.1 Dwarf galaxies

Figure 8 shows the results from the MHD_10 and MHD_SAT low-metallicity dwarf galaxy simulations of W23, which used AREPO. MHD_10 starts with an initially weak poloidal magnetic seed field of $10^{-6}\,\mu$G, while MHD_SAT has an initial poloidal field of $10^{-2}\,\mu$G. The data are taken from the 1 Gyr snapshots. Figure 9 shows the results from two new, high-resolution simulations that use the MHD_10 and MHD_SAT 1 Gyr data from W23 as initial conditions. We describe the new simulations in detail in Appendix B.

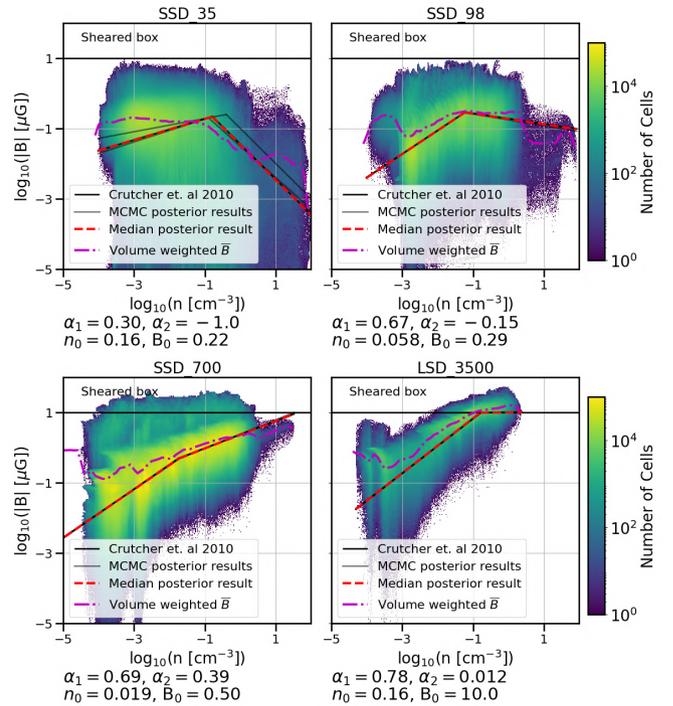

**Figure 7.** The volume-weighted absolute magnetic field $|\mathbf{B}|$ and number density $n$ distribution of PENCIL simulations of galactic dynamos. We use the notation introduced in Figure 4. We show three plots where the SSD is dominant at different times from Gent et al. (2024), and one where the LSD has become dominant. We see the effects of the growth of the SSD on the gas from its early kinematic stage at 35 Myr to when it is fully saturated at 700 Myr. We can see at 35 Myr has a decreasing field strength in dense gas. After 98 Myr we see a rise in the exponent of the slope in the denser gas to near $|\mathbf{B}| = n^0$ and in the diffuse gas it becomes larger. After 700 Myr when the SSD and field have had time to grow and reach saturation, we see a rise in the dense gas exponent to $\alpha_2 \sim 0.39$ whilst the diffuse region remains the same. The break occurs in this simulation towards the denser gas near where there are less cells. After 3500 Myr, the LSD is dominant and we see in this case a power law of $\alpha_1 \sim 0.78$, with a turn over and flattening in the most dense gas.

MHD_10 has both exponents nearly equal, $\alpha_1 = 0.49$ and $\alpha_2 = 0.44$ with the break density occurring at $n_0 \sim 360\,\text{cm}^{-3}$. The higher initial field model MHD_SAT has lower diffuse and dense gas exponents $\alpha_1 \sim 0.33$ and $\alpha_2 \sim 0.39$.

The higher resolution dwarf galaxy model MHD_10_HR shows similar exponents to MHD_10, $\alpha_1 = 0.47$ and $\alpha_2 = 0.49$ but with the break density occurring at much higher values; $n_0 \sim 700\,\text{cm}^{-3}$. Comparison of MHD_SAT_HR to MHD_SAT shows that the higher resolution simulation has a smaller exponent in the diffuse gas $\alpha_1 \sim 0.28$, but a higher exponent in the dense gas, reaching the observed range, $\alpha_2 \sim 0.54$.

#### 4.3.2 Milky Way-type galaxies

Figure 10 shows the results for a Milky Way-like galactic MHD simulation from Robinson & Wadsley (2024) using RAMSES for AGORA disk galaxy initial conditions (Kim et al. 2016) including a live dark matter halo. The data is taken after 600 Myr and has a resolution on an AMR grid ranging from 36.60 pc to 9.15 pc in the densest gas. The initial magnetic field was toroidal and set at either 1 $\mu$G or 0.1 $\mu$G at a number density of $n = 0.25$ cm$^{-3}$. The initial power-law index of the field-density relationship was set to $\alpha = 2/3$ across the whole density range and includes self-gravity. These simulations evolve the galaxy from this initial state over a period of 250–300 Myr, by which time the disk has settled into a steady state with a star formation rate of a few solar masses per year, as observed. The field saturates to 10–20 $\mu$G by the end of the simulation.

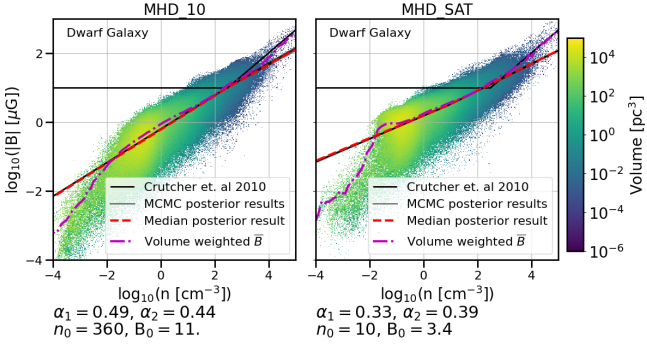

**Figure 8.** The volume-weighted absolute magnetic field |**B**| and number density $n$ distribution for dwarf galaxy simulations *(left)* MHD_10 and *(right)* MHD_SAT taken from W23. The same notation is used as Figure 4. Simulation MHD_10 shows an almost uniform power-law with $\alpha_1 \sim 0.49$ and a break at the expected $n_0 \sim 360$ cm$^{-3}$, whereas MHD_SAT has lower exponents in both the diffuse and dense gas with the break occurring at much higher density.

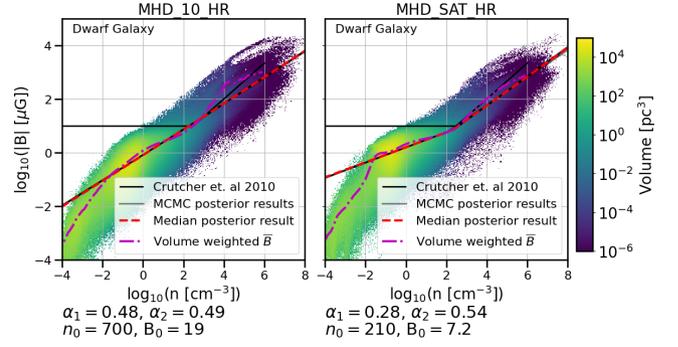

**Figure 9.** The volume-weighted absolute magnetic field |**B**| and number density $n$ distribution for the new high-resolution MHD dwarf galaxy simulations *(left)* MHD_10_HR and *(right)* MHD_SAT_HR. The same notation is used as Figure 4. MHD_10_HR again shows an almost uniform power-law exponent with $\alpha \sim 0.48$ with the break at $n_0 \sim 700$ cm$^{-3}$. MHD_SAT_HR shows a break at $n_0 \sim 210$ cm$^{-3}$ with a small exponent in the diffuse gas, and an exponent of $\alpha_2 \sim 0.54$ in the dense gas.

The exponents in both simulations have lower than the C10 values for $n_0$: 22 and 0.40 cm$^{-3}$ respectively. Both have large values for the exponent in the diffuse gas; MW_RAM_1 has $\alpha_1 \sim 0.56$ and MW_RAM_2 has $\alpha_1 \sim 0.72$. The dense gas exponent for MW_RAM_1 is $\alpha_2 \sim 0.14$ and for MW_RAM_2 is $\alpha_2 \sim 0.46$. The low value in MW_RAM_1 arises from the high density at which the break occurs and follows the volume-weighted field strength as expected.

Figure 11 shows two 3 kpc zoom-in regions of multi-scale galactic MHD simulations, using RAMSES and including SN feedback (Zhao et al. 2024). These have grid resolution of 0.286 pc with a mass refinement of 93 M$_\odot$ and are zoom-in boxes in a Milky Way-like simulation whose initial conditions are similar to Robinson & Wadsley (2024). The simulations employ delayed cooling that allows superbubbles to expand for 5 Myr before being affected by cooling. The zooms focus on two different 3 kpc regions within the galaxy: the *active* region is dominated by intersecting SN-blown bubbles at a time of 283 Myr (MWR_R2) while the *quiet* region (MWR_R1) is less affected by bubbles and follows a portion of a spiral arm at 337 Myr. The dense gas is resolved to $n = 10^6$ cm$^{-3}$.

The exponents vary between the two different simulation regimes. For the dense gas exponents, the quiet region MWR_R1 has $\alpha_2 \sim 0.52$ while the active region MWR_R2 has $\alpha_2 \sim 0.44$. The more turbulent active region has a slightly lower value. For the diffuse gas exponent, on the other hand, there is the opposite trend: the quiet region MWR_R1 has $\alpha_1 \sim 0.59$, while the active region MWR_R2 shows $\alpha_1 \sim 0.61$.

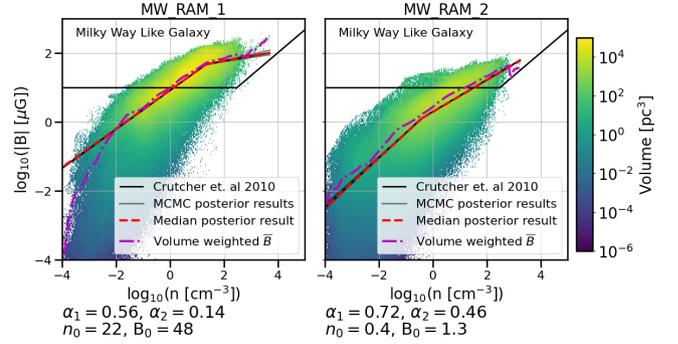

**Figure 10.** The volume-weighted absolute magnetic field |**B**| and number density $n$ distribution taken from a Milky Way simulation using RAMSES from Robinson & Wadsley (2024). The same notation is used as Figure 4. *(Left)* a simulation with a higher initial field strength of $1\mu G$. *(Right)* a simulation with a lower initial field strength of $0.1\mu G$. Both simulations show a higher value of $\alpha_1$ than $\alpha_2$, with MW_RAM_2 being the larger. MW_RAM_1 shows a large decrease in $\alpha_2$, but with a higher break density and field strength.

## 5 DISCUSSION

### 5.1 The $B$-$n$ relation from observations

The original $B$-$n$ relation has been a useful tool in furthering our understanding of the physics of magnetic fields in the ISM, with its implications for star formation and other processes. However, since it was derived by C10 from their Zeeman analysis, there has been little development of the approach beyond the use of Zeeman measurements (see Tritsis et al. 2015; Jiang et al. 2020), and one study fitting a single power law to DCF data (Liu et al. 2022b).

With our extended analysis of not only the Zeeman data, but also of 24 years of starlight and dust polarized emission observations and DCF calculations, we have found that a new form for the relation emerges, which has important consequences.

We begin our discussion by considering our treatment of the original Zeeman data and how it differs from the earlier

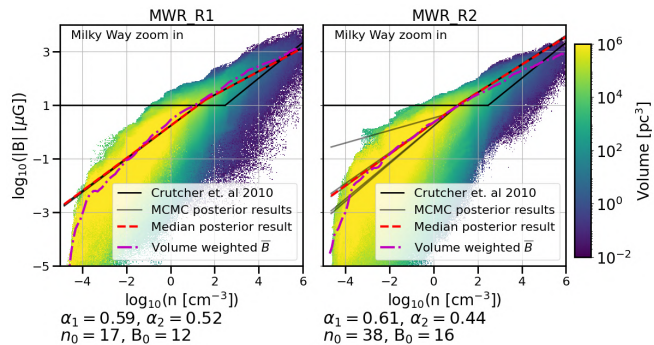

**Figure 11.** The volume-weighted absolute magnetic field |**B**| and number density $n$ distribution taken from two, 3 kpc zoom-in regions of Milky Way simulations from Zhao et al. (2024) using RAMSES: *(Left)* an active region dominated by superbubbles, *(Right)* a more quiescent region dominated by a spiral arm segment. The same notation is used as Figure 4. We see that the maximum a posteriori relation has a larger $\alpha_1$ for both regions ($\sim 0.59$ and 0.61 respectively) than $\alpha_2$

work. The power-law exponents $\alpha_1$ and $\alpha_2$ that we derive are inconsistent with the values proposed in C10. Our high-density exponent $\alpha_2 = 0.53$, below the original 0.66, which lies just outside our 68% credible interval. Our low-density exponent $\alpha_1$ appears to be well constrained with the 68% interval being relatively small. We also find that our median value for $n_0 \sim 4000$ cm$^{-3}$ is much higher than C10, whose value does not lie within our 68% interval.

We also fit the DCF data set with the same two-part power law as for the Zeeman data, resulting in a lower value of the diffuse gas exponent $\alpha_1 \sim 0.26 < \alpha_2$. Neither of these results match the $\alpha \sim 0.57$ found in the analysis of DCF data by Liu et al. (2022b), though they only consider a single power-law using a least-squares fit. Much like in our analysis of the Zeeman data, the DCF analysis shows the break occurring at a higher density and field strength than C10, though both with significant uncertainties.

What determines the exponents? Sur et al. (2012) argue that different turbulent driving processes can affect the exponent in the dense gas. For example, SNe, shear, and gravitational collapse all affect the exponents. Another possibility is whether the gas is magnetically super-critical or not (Li et al. 2013; Mocz et al. 2017; Myers & Basu 2021; Auddy et al. 2022). The higher the magnetic field, the less supercritical the gas, the lower the diffuse gas exponent, and the sharper the difference between the two exponents becomes (Mocz et al. 2017).

Our most interesting finding is the similarity of both exponents in the two data sets, with both the median and MAP exponents derived from the Zeeman measurements being close to the DCF results, showing that the DCF data can be used to derive a relationship despite its issues.

A more detailed study of the diffuse exponent, including detailed error treatment, would provide valuable insights. Its value likely probes the dynamical nature of magnetic field generation in diffuse gas. We shall discuss this in more detail in the context of numerical simulations in Section 5.3.

The interpretation of a slope change at the density $n_0 \sim 300$ cm$^{-3}$ given by C10 is that it is the average density at which parsec-scale clouds become self-gravitating. However, this is not a strict threshold, and the break can occur across a range of values, as our results show. Our results suggest that the break in the *B*-*n* relation depends on a number of physical processes, including not only gravitational collapse, but also flow convergence, SN-driven shock fronts, and other forms of turbulence. We discuss the break in detail in Section 5.2.

One concern in using the DCF as a complementary method to Zeeman measurements is that the DCF method is well known to over-predict the field strength giving values that can be up to $\sim 6$ times higher than Zeeman data (Pattle et al. 2023; Hwang et al. 2024). Heitsch et al. (2001) performed synthetic dust polarisation observations (DCF-like observations) on the data produced by simulations of three-dimensional MHD turbulence driven for self-gravitating molecular clouds. They derive an increase in field strength by a factor of a few from the synthetic observations, similar to what is expected to occur in the ISM, so they argued that the DCF method is viable for observations. Chen et al. (2022) also model synthetic DCF observations as well as structure function-DCF observations, noting that this is a preferred method. Again, their synthetic results are similar to those from observations, but uncertainties in the linewidth may be more important than errors in polarization. However, Liu et al. (2022b) produce a relation similar to our Zeeman results using a least-squared fit to their DCF data, as discussed above.

These studies suggest that using the DCF data is valid as long as we acknowledge its limitations. Crucially, by analysing both the Zeeman and DCF data sets we see that the diffuse gas exponent is roughly consistent with the DCF result covering seven orders of magnitude. However, the dense gas exponents are not consistent. This likely arises from the lack of data in the Zeeman observations above $\sim 10^4$ cm$^{-3}$ and likely inaccuracies in the DCF observations at high densities. Despite this and the method's known tendency to overestimate magnetic field strength, it can effectively capture a scaling relationship between magnetic field and density.

### 5.2 The observational break density

One of the most notable differences between our work and previous studies is the inference of a new free parameter $\alpha_1$ for the power-law of the relationship in diffuse gas. Adding $\alpha_1$ changes key aspects of the original relationship, most importantly, the value of the break density $n_0$.

Figure 12 shows the marginal distributions for $n_0$ from our hierarchical Bayesian analyses for the Zeeman and DCF data set. The estimated $n_0$ for both is significantly different from that predicted by C10. The Zeeman data implies $n_0 \sim 4000$ cm$^{-3}$, a full decade higher than the C10 result. One reason for this could be the way the Zeeman data are distributed. Recall that the Zeeman data are derived using the line emission from three distinct species that sample different density ranges. One problem that arises from this is that there are two distinct gaps in the data that separate the three surveys, one at $n \sim 10^3$ cm$^{-3}$ and one at $n \sim 5 \times 10^4$ cm$^{-3}$. A result of this is that 100 of the 137 observations lie below $n = 4000$ cm$^{-3}$. This skew to lower densities could bias the fit due to clumpy, non-randomly sampled distribution of density measurements. These variations in the distribution of the Zeeman data over $n$ could then contribute to the uncertainty in the break density. To better constrain the break in the Zeeman data we need better sampling of the density range.

This leads to another issue with the Zeeman data: our lack of knowledge of the errors for the number densities. We have assumed a factor of five in our analysis which allows for a large range of possible values for $n_0$. We can see this in the wide distribution of $n_0$ in Figure 12. Appendix A3.4 examines different factors for $n_0$ and discusses our choice in more detail.

The DCF data implies $n_0 \sim 1.40 \times 10^5$ cm$^{-3}$, a break density much higher than that inferred from the Zeeman value. The DCF data also have a much higher value of $n_0$ than the original Zeeman results reported in C10, which lies far outside our 68% credible interval. The distribution of the DCF data is quite even across decades of $n$ and therefore the inferred break density from the DCF data is likely more trustworthy. What could be driving this high $n_0$ value?

The ISM is a multiphase environment impacted by many different physical processes. Interestingly, densities of $n_0 \sim 10^5$ are the point at which dense star-forming cores begin to form (Bergin & Tafalla 2007; Enoch et al. 2007; Pattle et al. 2023), so the gas is starting rapid collapse. Most observations of DCF at this density and higher are in dense cores. It has been shown recently (Ibáñez-Mejía et al. 2022; McGuiness et al. 2025), that at high densities magnetic energy becomes less dominant and the kinetic energy becomes more important. This reduction in the relative importance of the field in supporting gas against collapse could be a reason for the break to occur at these high densities.

There are observational examples of regions where stellar feedback produces a *B-n* scaling that is above the reported trends. Troland et al. (2016) report field strengths 3–5 times above the mean at comparable densities toward the atomic photon-dominated region that lies just in front of the Trapezium stars of Orion. Joubaud, T. et al. (2019) estimate enhanced *B*-field strengths using DCF toward the lower rim of the Orion-Eridanus superbubble. Due to the scarcity of observational data, these particular local conditions cannot be easily marginalized to identify a single physical cause or a value for the break density. Soler & Hennebelle (2017) show that a negative velocity divergence can cause a change in field orientation in observations at around the original break density and conclude that this could arise for multiple reasons, i.e. gravitational collapse, external pressure from the warm medium, or colliding flows. These degeneracies could also lead to changes in the break density. These results suggest that the *B-n* scaling is strongly tied to the specific environment, and if so, more extended surveys of field strengths will be needed to draw further conclusions on the generality of the critical density.

### 5.3 The *B-n* relation from numerical simulations

In this subsection, we discuss the *B-n* relations drawn from our various simulations and reasons for the deviations seen in each case.

#### 5.3.1 Core collapse simulations

These simulations are of sufficiently high resolution to capture the formation of protostellar discs, which are much denser than typical regions in molecular clouds, as we see in Figure 4, where $n = 10^{13}$ cm$^{-3}$ is reached. The negative $\alpha_1$ is due to numerical scatter in the most diffuse gas. These simulations do not trace gas below $\sim 500$ cm$^{-3}$. This would technically prevent us from drawing conclusions on the diffuse gas exponent and break density suggested by C10. However, comparing with our new observational results we do not find a break at higher densities. This could be due to the result of focusing on discs in the model.

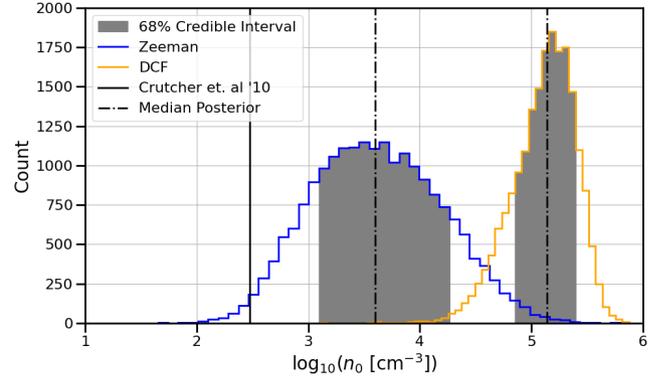

**Figure 12.** Marginal posterior distributions for $n_0$ for the Zeeman *(blue)* and DCF *(orange)* data sets. The shaded regions show the 68% credible intervals, the dashed-dotted lines show the posterior medians, and the solid black line shows the prediction from C10. Neither distribution favours the latter result.

Protostellar discs are born in an environment undergoing compression through gravitational collapse. This gives rise to $\alpha_2 \sim 0.56$ as the gas begins to collapse perpendicular to the field, compressing the field lines. We see no difference in the value between simulations with and without ambipolar diffusion, although we note the presence of a plateau at high density (in a negligible fraction of the volume) for the case with ambipolar diffusion.

#### 5.3.2 Stratified Box simulations

The two $(0.5 \text{ kpc})^3$ stratified box simulations of Girichidis et al. (2018) shown in Figure 5 only differ in initial seed field strength. They reach high enough densities that a break at $n_0 = 300$ cm$^{-3}$ could appear, but it actually appears at lower values of, $n_0 \sim 0.45$ cm$^{-3}$ and $n_0 \sim 0.26$ cm$^{-3}$. The diffuse exponent is high whilst the dense exponent is much lower than expected.

There are two possible reasons for this result. The first is that the diffuse gas is more turbulent than the dense gas as the simulation is being driven by strong SN feedback. Gent et al. (2021) report that the SSD grows rapidly in hot, diffuse gas, so the high sound speeds provided by SN shock heating accelerate field growth. This suggests that not enough SN heating has occurred to allow for growth of the SSD. The second explanation is that these simulations have run for only 60 Myr, meaning the dynamo may not have reached saturation and the field is still growing. The SSD can take from a few to hundreds of megayears to saturate depending on numerical resolution (Gent et al. 2021; Whitworth et al. 2023; Gent et al. 2024).

Unsaturated fields in these simulations have important effects. With values significantly below saturation, the magnetic pressure support is reduced, allowing the gas to collapse faster. Starting with a higher field (simulation B6, Fig. 5) does not seem to counteract this. Figure 3 of Robinson & Wadsley (2024) shows their high initial field strength simulation decreasing in field strength over time, meaning an artificially high field may not be able to counter act this effect and that allowing a field to evolve naturally, or begin near saturation is a better solution.

The simulations in Figure 6 have been run for a longer time than those in Figure 5. Both show small exponents in both the diffuse and dense gas, which means that evolutionary time could have an effect. The gas is hot and turbulent enough for the SSD to grow. However, when the resolution is increased in KPC_HIGH and the simulation is run longer, we see a smaller exponent in the diffuse gas.

Again, there are a few reasons why this could be happening. One is the change in resolution, as discussed in Section 5.3.4. Since the saturation time for the SSD depends sensitively on numerical resolution down to sub-parsec scales (Gent et al. 2021), a sudden change in resolution will lead to a change in the rate of field amplification. As KPC_HIGH was only run for an additional $\sim 20\,\mathrm{Myr}$ at higher resolution, this may not be long enough for the B field to saturate at the new resolution, leaving the field strength low at high densities.

Another key factor in determining the exponents are the initial conditions. Sur et al. (2012) showed that varying the turbulent driving can lead to a departure from the dense gas exponent $\alpha_2 = 2/3$. It becomes smaller as the SSD saturates first on small scales, and later at larger scales. Gent et al. (2021) show that the SSD grows rapidly in a hot turbulent medium. Most simulations start with the gas uniform in temperature and density. Therefore, if simulations are initiated with specific seed field configurations, then the magnetic field in the diffuse medium will likely drop due to the lack of a hot, turbulent phase.

A possible cause for the exponents found in the SSD and LSD simulations that specifically lack self-gravity (Fig. 7) is the equipartition between the magnetic and kinetic energies. This would introduce an additional variable into the relationship, the dependence on gas velocity, changing the relationship from $B \propto n^\alpha$ to $B \propto vn^\alpha$. However, testing this new form is beyond the scope of this work.

### 5.3.3 Galactic rotation

The dwarf galaxy simulations of W23, shown in Figure 8, have a range in density of $n = 10^{-4}$–$10^5\,\mathrm{cm}^{-3}$, comparable to that covered by the Zeeman observations. We see that MHD_SAT has smaller exponents in both diffuse and dense gas than MHD_10, though in both cases the diffuse and dense exponents are similar. This suggests that the relationship is connected to the initial conditions. We confirm this by noting that MHD_SAT has a larger initial seed field, $10^{-2}\,\mu\mathrm{G}$, which, as seen in Figure 5 of W23, results in a field with large-scale order and higher strength in the diffuse regions.

If a simulation is initiated with a stronger field, then the gas is dynamically supported for longer. This leads to the diffuse gas starting with a higher field, giving rise to a lower value for $\alpha_1$. This is in contrast to the stratified box simulations and is likely an effect of numerical resolution, with the stratified box simulations having rather higher resolution, allowing them to model faster growth of fields from small scales to saturation at large scales.

Figure 10 shows the results from a Milky Way-scale simulation. As in the case of the dwarf galaxy simulations, the diffuse gas field strength has dropped below the initial field strength. This suggests the SSD is not active enough to amplify the field here. This may be a resolution problem, as the diffuse gas is under-resolved, meaning the SSD cannot be accurately modelled (see Martin-Alvarez et al. 2022, for discussion of this issue). We discuss this in Section 5.3.4 below.

Figure 2 in Robinson & Wadsley (2024) compares the time evolution of the $B$-$n$ relationship for different initial field strengths. In the case of a weak or medium initial seed field, $\alpha_2$ tends to 0.5. In the strong field case, MW_RAM_1, the diffuse gas exponent rises to $\alpha_1 = 0.56$ but $\alpha_2$ drops significantly to 0.14. However, in the weak field case both $\alpha_1$ and $\alpha_2$ grow from the initial conditions. The important point here is that simulations with weak initial fields grow naturally by dynamo action resulting in the closest matches to the observations. Imposing artificially strong initial fields has adverse effects on star formation and the final saturated state of the field.

Large-scale dynamo action in Milky Way-scale galaxies depends on rotational time scales, which are much longer than the eddy time on which the SSD grows. It is expected to take of the order of 5–10 rotations (Pfrommer et al. 2022; Gent et al. 2024) for the LSD to become the dominant source of magnetic field support. The rotational period of the dwarf galaxies is on the order of 400 Myr (W23), with a similar dynamical time in the simulations of Robinson & Wadsley (2024). As these simulations only run for 1 Gyr, this likely means that LSD effects are not yet dominant.

The simulations in Figure 7 by Gent et al. (2021) and Gent et al. (2024) track the effects of dynamo evolution over 3.5 Gyr. We see that as the dynamo evolves from the SSD in the early stages ($\sim 35$ Myr) to the LSD at late times ($\sim 3500$ Myr), the diffuse exponent characterizing the bulk of the gas tends to high values, $\alpha_1 \sim 0.78$. This could point to the denser and more turbulent gas in these models amplifying the field quicker. Although we cannot say much about the dense gas exponent from these simulations, as they do not trace a density range high enough, we can be confident in saying that the diffuse gas has a non-zero exponent and that this arises purely from dynamo activity; these simulations do not include self-gravity, implying that their $B$-$n$ relation is purely driven by dynamo activity driven, feedback and shear.

In summary, we have shown that all simulations that include large scale galactic rotation tend towards an exponent of $\geq 0.5$. This can appear in either the diffuse gas exponent $\alpha_1$ or the dense gas exponent $\alpha_2$ in the results, depending on the setup of the simulation and the captured density range. We also see that the dwarf galaxy simulations and the Milky Way-like simulations are most similar to the observational results. This is one of our most important results. The broadest view of the simulations then is that in order to produce a relationship similar to the observed $B$-$n$ relationship, they should include both dense and diffuse gas, self-gravity, along with shear and feedback.

### 5.3.4 Resolution effects on the dynamo

Using high resolution zoom-in MHD simulations of clouds embedded in a stratified box with SN driving, Ibáñez-Mejía et al. (2022) showed that molecular clouds have magnetically supported envelopes with a field perpendicular to the density gradient. In all three clouds they simulate, each has an exponent of less than $\alpha = 0.5$ with no break in the *B*-*n* relationship. However, the diffuse gas is poorly resolved in these AMR simulations.

Indeed, most numerical simulations of the ISM on large scales lack resolution in diffuse gas. Of the simulations presented here, two resolve the diffuse gas to scales where the SSD should be captured: the dynamo simulation of Gent et al. (2024), SSD_35, SS_98, and SSD_700, with a uniform 1 pc resolution, and the Milky Way-like zoom-in simulation of Zhao et al. (2024), with a uniform 0.29 pc resolution. This is important as the SSD must be resolved down to 1 pc or better scales to accurately capture its growth rate (Korpi-Lagg et al. 2024). Resolving the SSD is important in order to track the rapid amplification intrinsic to turbulent dynamo theories. This could lead to a stronger field in the diffuse gas adding support and slowing its transport to denser regions.

In the time-evolving SSD PENCIL simulation (Fig. 7), $\alpha_1$ tends towards a larger than expected value of 0.78. This is a stratified box simulation with shear, not a full galactic disc, but it does have a dense mid-plane with diffuse gas bubbles (see Figures 1, 3, and 13 in Gent et al. 2024). This shows that given enough time, on the order of a few hundred megayears, in a simulation with a rotation rate twice that of the Milky Way (so around five rotation periods), along with shear, an exponent exceeding 0.5 in the diffuse gas can arise.

The high resolution dwarf galaxy simulations MHD_10_HR and MHD_SAT_HR (Fig. 9) have little difference in the low density exponents compared to the original simulations of W23 (Fig. 8). This is due to the diffuse gas still not being well resolved, only reaching parsec scale resolution at $n > 10$ cm$^{-3}$. However, in the dense gas for MHD_SAT_HR we find the exponent has increased, matching the observed Zeeman result. This model includes photoionisation feedback and high resolution at the highest densities and has had an ordered LSD for longer than other models. This shows that the more complete the model, the more likely we are to approach the observed results. When we look at the lowest resolution dwarf galaxy simulation, MHD_10_LR, we see a large departure from the results in the more resolved simulations. Appendix C examines the difference between the high resolution and base resolution dwarf galaxy simulations, specifically in the diffuse gas.

The zoom in regions of Zhao et al. (2024) shown in Figure 11 have a similar exponent in both the dense and diffuse gas. However, the turbulent region, MWR_R2, has a larger exponent in the diffuse gas, which supports the idea that turbulence is driving the relationship.

Thus, we have found that the simulations that match our observational results the closest are, primarily the high-resolution dwarf galaxy simulation MHD_SAT_HR, followed by MHD_10_HR and MWR_R1. These are all high resolution simulations best able to trace both diffuse and dense gas, meaning they are more able to accurately model both the small and large scale dynamos.

In summary, by considering the ensemble of simulations presented here, a major finding is that the more physically complete a simulation is, the more closely the exponents in the relationship match the observed exponents. In addition, this match is better when simulations are allowed to evolve for a sufficiently long time, when they spatially resolve both the diffuse and dense phases, and when they include galactic rotation or shear as well as stellar feedback mechanisms.

### 5.4 The numerical break density

Only a few of the simulations presented here allow us to confidently draw conclusions about the break density $n_0$, due to the density ranges they cover. Simulations that do not span a density range a few orders of magnitude above and below $n \simeq 100$ cm$^{-3}$ will likely give misleading results. For example, in the protoplanetary disc population simulations NI and IMHD, we find $n_0 \sim 700$ cm$^{-3}$ which is close to their lowest density. Conversely, the PENCIL simulations don't reach densities $n_0 > 300$ cm$^{-3}$ meaning no break is likely to be found at the expected density.

In summary, we find a broad range of values of the break density for the simulations that span the density range of the observational data sets. These are the simulations KPC_HIGH, MHD_10, MHD_SAT, MHD_10_HR, MHD_SAT_HR, MWR_R1, MWR_R2, and possibly B3 and B6. The lowest $n_0$ found in these simulations is $n_0 \simeq 17$ cm$^{-3}$ in MWR_R2, whilst the highest is $n_0 \simeq 700$ cm$^{-3}$ in MHD_10_HR—a difference of more than an order of magnitude.

Of these simulations, only the top of the 68% credible interval of MHD_10_HR comes close to the lower end of the Zeeman-derived 68% credible interval, whilst none of the simulations reach the lower end of the DCF 68% credible interval. MHD_SAT_HR and MHD_10 give results close to the C10 break density of $n_0 = 300$ cm$^{-3}$. (The values for $B_0$ are just as variable with numbers ranging from $B_0 \sim 0.094 \mu$G in MHD_10_LR to $B_0 \sim 48 \mu$G in MW_RAM_1.)

The numerical break density appears to depend on several key factors, including numerical resolution, initial seed field strength, and the run time of the simulation. To better constrain the break point in numerical simulations, we need better estimates on the errors. We may also need to resolve the dense gas phase more carefully. If, for example, the reason for the break in the DCF data is dense core formation, then this would need to be fully resolved with accurate feedback models that include winds and jets generated by the protostellar objects. We leave this for future studies.

### 5.5 Numerical caveats

A consideration for any numerical simulation is the effect of numerical diffusion as a result of finite resolution. In magnetic field studies this is particularly important because of the effects at small scales where resistivity becomes important. None of the simulations presented here are resolved well enough to completely overcome this. We can, however, consider what happens when resolution is improved (Fig. 13). MHD_10_LR is taken from the Appendix of W23 and has a much lower resolution than MHD_10 with a Jeans refinement of 4 cells instead of 8, a cell mass of 500 M$_\odot$, and a sink creation density threshold of $10^2$ cm$^{-3}$. Whilst the dense gas exponent is $\alpha_2 \sim 0.52$, the diffuse exponent becomes much larger, $\alpha_1 \sim 0.84$. As we increase the resolution up to the high resolution simulation MHD_10_HR, $\alpha_1$ tends to $\sim 0.49$.

Another caveat is that the simulation results do not typically have errors. We have attempted to rectify this by implementing errors based on the KDE method as described in Appendix A3. Comparing the results of from Bayesian inference to the volume weighted average magnetic field binned via density in each numerical simulation in Figures 4–11, we see that they match quite closely. By including the nuisance factor $\ln V$ in the model we mitigate some of the issues that arise due to data without well-defined errors.

## 6 SUMMARY AND CONCLUSIONS

We have comprehensively treated the observed $B$–$n$ relation by analysing two observational data sets—the original Zeeman data of C10 and a large sample of DCF measurements—within a hierarchical Bayesian framework that allows both exponents in the power-law relation to vary freely.

Our results differ from those found by C10. This is most clearly seen in the non-zero power-law exponent for the diffuse gas as well as in the values of the break density $n_0$ and magnetic field $B_0$. The break density in the DCF results is roughly comparable to that of dense star forming cores, $n_0 \simeq 10^5$ cm$^{-3}$. However, the ISM is highly anisotropic and without further more detailed study the exact reason for the break value is difficult to determine.

We have also performed a Bayesian analysis on an ensemble of nineteen numerical simulations, resulting in a wide range of parameters for the $B$-$n$ relation. The simulation results are highly dependent on the system being studied. For simulations to reproduce the observed results, a turbulent galactic dynamo must resolve both the diffuse and dense gas (requiring a spatial resolution below 1–2 pc) and be run over sufficiently long time scales (a few hundred megayears) to bring an initially weak magnetic field up to a saturated, steady-state value in the ISM. Thus, it appears that the relation in diffuse gas is not strongly tied to the self-gravity of the system. Rather, it is driven by the dynamo, feedback, and turbulence.

We summarise our key findings:

• We fit a two-part power-law relation differing from that found by C10 for the observational data(Eq. 2).
• The parameters for the Zeeman data are $B_0 = 95.00^{+60.00}_{-36.00}\mu$G, $n_0 = 0.400^{+1.30}_{-0.30} \times 10^4$ cm$^{-3}$, $\alpha_1 = 0.15^{+0.06}_{-0.09}$, and $\alpha_2 = 0.53^{+0.09}_{-0.07}$, though $\alpha_2$ is strongly dependent on the errors assumed for $n$.
• The DCF data that has reported errors in Pattle et al. (2023) results in parameters $B_0 = 460.00^{+120.00}_{-100.00}\mu$G, $n_0 = 14.00^{+10.00}_{-7.00} \times 10^4$ cm$^{-3}$, $\alpha_1 = 0.26^{+0.15}_{-0.15}$, and $\alpha_2 = 0.77^{+0.14}_{-0.15}$.
• The observed exponent in the diffuse gas $\alpha_1 > 0$ in all cases.
• The observed break density occurs at substantially higher densities than found by C10, although with large error bars reflecting the highly variable dynamical and anisotropic state of the ISM.
• The high-resolution dwarf galaxy simulations match the observed power-law exponents most closely, while the other full galaxy simulations also give good results.
• We find clear evidence that the less approximate the simulations are the more likely they are to match the observed $B$-$n$ relation. Important improvements include higher spatial resolution, inclusion of shear and other dynamical processes, and long enough run times to accommodate dynamo saturation and other turbulence-related effects.

To make further progress, we require more data from both observations and numerical simulations. It is exciting that Zeeman observations (Thompson et al. 2019; Hwang et al. 2024), Faraday rotation, and synchrotron polarization observations in the epoch of the Square Kilometer Array will offer new insights into the diffuse gas. The rapidly increasing dynamical range of multiscale galactic MHD simulations is also promising. We shall examine these in detail in future works.

## ACKNOWLEDGEMENTS

We thank the members of the AREPO ISM group and ECO-GAL Large Scale Structures group for discussions and insightful comments on the coding and science goals in this paper. The anonymous referee provided a very useful report that helped us to improve the manuscript in many ways. We also thank V. Springel for access to AREPO and R. Treß for the development of the photoionization code used in this work. DJW acknowledges support from the Programa de Becas Posdoctorales of the Dirección General de Asuntos del Personal Académico of the Universidad Nacional Autónoma de México (DGAPA,UNAM,Mexico). SS acknowledges support from UNAM-PAPIIT Program IA10484. REP is supported by a Discovery Grant from NSERC, Canada. He is also grateful for support of his sabbatical leave in Heidelberg (2022/23), by the Institute for Theoretical Astrophysics of the University of Heidelberg (ITA) and the Max Planck Institute for Astronomy, where this work was started. RJS gratefully acknowledges an STFC Ernest Rutherford fellowship (grant ST/N00485X/1). M-MML acknowledges direct support from US National Science Foundation grants AST18-15461 and AST23-07950, as well as support from grant PHY23-09135 to the Kavli Institute for Theoretical Physics. LSV was supported by the Science Pathways Scholars Program of Barnard College. RP acknowledges funding support from Ontario Graduate Scholarships and the Queen Elizabeth II Graduate Scholarship in Science & Technology. AP acknowledges financial support from the grant UNAMPAPIIT IG100223, the Sistema Nacional de Investigadores of CONAHCyT, and from the CONAHCyT project number 86372 of the 'Ciencia de Frontera 2019' program, entitled 'Citlalcóatl: A multiscale study at the new frontier of the formation and early evolution of stars and planetary systems', México.

DJW, PG, NB, UL, PH, RSK, and JDS acknowledge funding from the European Research Council via the ERC Synergy Grant "ECOGAL" (project ID 855130). RSK is grateful to the German Excellence Strategy for funding via the Heidelberg Cluster of Excellence "STRUCTURES" (EXC 2181-390900948), and to BMWK for support in project "MAINN" (funding ID 50OO2206). The team in Heidelberg acknowledges computing resources provided by *The Län* and DFG through grants INST 35/1134-1 FUGG and 35/1597-1 FUGG, data storage at SDS@hd funded through grants INST 35/1314-1 FUGG and INST 35/1503-1 FUGG, as well as computing resources provided by the Leibniz Rechenzentrum via grants pr32lo, pr73fi and GCS large-scale project 10391. RSK thanks the 2024/25 Class of Radcliffe Fellows for highly interesting and stimulating discussions.

This work used the DiRAC COSMA Durham facility managed by the Institute for Computational Cosmology on behalf of the STFC DiRAC HPC Facility (www.dirac.ac.uk).

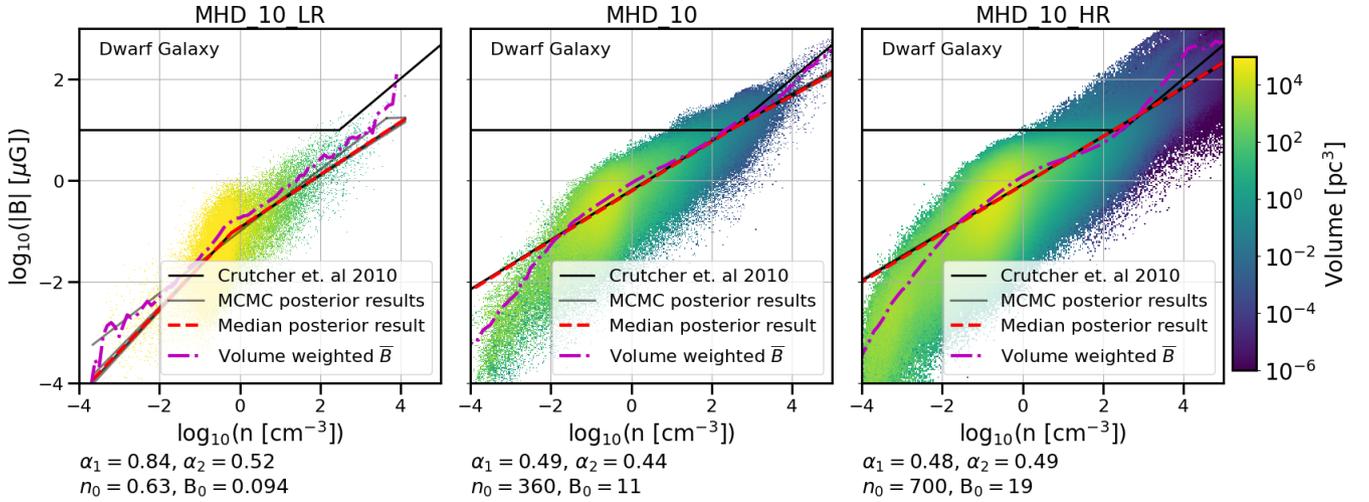

**Figure 13.** The |**B**|-$n$ distribution for three dwarf galaxy simulations with varying numerical resolution: the low resolution dwarf galaxy simulation from the Appendix of W23, MHD_Low, MHD_10, and the new high-resolution MHD_10_HR. The same notation is used as Figure 4. This comparison shows the effects of numerical diffusion. As the simulation becomes more resolved, the relationship for both exponents tends to $\alpha_1 \simeq \alpha_2 \simeq 0.5$.

The equipment was funded by BEIS capital funding via STFC capital grants ST/P002293/1, ST/R002371/1 and ST/S002502/1, Durham University and STFC operations grant ST/R000832/1. DiRAC is part of the National e-Infrastructure. The research conducted in this paper used SciPy (Virtanen et al. 2020), NumPy (van der Walt et al. 2011), and matplotlib (Hunter 2007).

## DATA AVAILABILITY

The dwarf galaxy data is available upon request to the first author. For all other data the reader will need to contact the owner of that data as listed in the appropriate reference.

## REFERENCES

Angarita Y., Versteeg M. J. F., Haverkorn M., Marchal A., Rodrigues C. V., Magalhães A. M., Santos-Lima R., Kawabata K. S., 2024, AJ, 168, 47
Auddy S., Basu S., Kudoh T., 2022, ApJ, 928, L2
Banerjee R., Vázquez-Semadeni E., Hennebelle P., Klessen R. S., 2009, MNRAS, 398, 1082
Beck R., 2015, A&ARv, 24, 4
Bergin E. A., Tafalla M., 2007, ARA&A, 45, 339
Brandenburg A., Dobler W., 2002, Computer Physics Communications, 147, 471
Brandenburg A., et al., 2020, Journal of Open Source Software, 6, 7
Brucy N., Hennebelle P., Colman T., Iteanu S., 2023, A&A, 675, A144
Cao Z., Li H.-b., 2023, ApJ, 946, L46
Carilli C. L., Taylor G. B., 2002, ARA&A, 40, 319
Chandrasekhar S., Fermi E., 1953, ApJ, 118, 113
Chen C.-Y., Li Z.-Y., Mazzei R. R., Park J., Fissel L. M., Chen M. C. Y., Klein R. I., Li P. S., 2022, MNRAS, 514, 1575
Colman T., et al., 2022, MNRAS, 514, 3670
Commerçon B., González M., Mignon-Risse R., Hennebelle P., Vaytet N., 2022, A&A, 658, A52
Crutcher R. M., 2004, in Uyaniker B., Reich W., Wielebinski R., eds, The Magnetized Interstellar Medium. pp 123–132
Crutcher R. M., 2012, ARA&A, 50, 29
Crutcher R. M., Nutter D. J., Ward-Thompson D., Kirk J. M., 2004, ApJ, 600, 279
Crutcher R. M., Wandelt B., Heiles C., Falgarone E., Troland T. H., 2010, ApJ, 725, 466
Davis L., 1951, Physical Review, 81, 890
Elia D., et al., 2017, MNRAS, 471, 100
Enoch M. L., Glenn J., Evans II N. J., Sargent A. I., Young K. E., Huard T. L., 2007, ApJ, 666, 982
Foreman-Mackey D., Hogg D. W., Lang D., Goodman J., 2013, PASP, 125, 306
Fryxell B., et al., 2000, ApJS, 131, 273
Gelman A., Rubin D. B., 1992, Statistical Science, 7, 457
Gent F. A., Mac Low M.-M., Käpylä M. J., Singh N. K., 2021, ApJ, 910, L15
Gent F. A., Mac Low M.-M., Korpi-Lagg M. J., 2024, ApJ, 961, 7
Girichidis P., Seifried D., Naab T., Peters T., Walch S., Wünsch R., Glover S. C. O., Klessen R. S., 2018, MNRAS, 480, 3511
Girma E., Teyssier R., 2023, arXiv e-prints, p. arXiv:2311.10826
Heitsch F., Zweibel E. G., Mac Low M.-M., Li P., Norman M. L., 2001, ApJ, 561, 800
Hennebelle P., Banerjee R., Vázquez-Semadeni E., Klessen R. S., Audit E., 2008, A&A, 486, L43
Hennebelle P., Commerçon B., Lee Y.-N., Charnoz S., 2020, A&A, 635, A67
Hildebrand R. H., Kirby L., Dotson J. L., Houde M., Vaillancourt J. E., 2009, ApJ, 696, 567
Houde M., Vaillancourt J. E., Hildebrand R. H., Chitsazzadeh S., Kirby L., 2009, ApJ, 706, 1504
Hull C. L. H., Zhang Q., 2019, Frontiers in Astronomy and Space Sciences, 6, 3
Hunter J. D., 2007, Computing in Science Engineering, 9, 90
Hwang J., Lee C. W., Kim J., Chung E. J., Kim K.-T., 2024, ApJ, 974, 231
Ibáñez-Mejía J. C., Mac Low M.-M., Klessen R. S., 2022, ApJ, 925, 196
Jiang H., Li H.-b., Fan X., 2020, ApJ, 890, 153
Joubaud, T. Grenier, I. A. Ballet, J. Soler, J. D. 2019, A&A, 631, A52
Kim J.-h., et al., 2016, The Astrophysical Journal, 833

Kirk H., Klassen M., Pudritz R., Pillsworth S., 2015, ApJ, 802, 75
Konstantinou A., Ntormousi E., Tassis K., Pallottini A., 2024, A&A, 686, A8
Könyves V., et al., 2015, A&A, 584, A91
Korpi-Lagg M. J., Mac Low M.-M., Gent F. A., 2024, Living Reviews in Computational Astrophysics, 10, 3
Kunz M. W., Mouschovias T. C., 2010, MNRAS, 408, 322
Lazarian A., 2020, ApJ, 902, 97
Lebreuilly U., Hennebelle P., Colman T., Commerçon B., Klessen R., Maury A., Molinari S., Testi L., 2021, ApJ, 917, L10
Li H.-b., Fang M., Henning T., Kainulainen J., 2013, MNRAS, 436, 3707
Li J., Jiang B., Zhao H., Chen X., Yang Y., 2024, ApJ, 965, 29
Liu J., et al., 2019, ApJ, 877, 43
Liu J., Zhang Q., Commerçon B., Valdivia V., Maury A., Qiu K., 2021, ApJ, 919, 79
Liu J., Zhang Q., Qiu K., 2022a, Frontiers in Astronomy and Space Sciences, 9, 943556
Liu J., Qiu K., Zhang Q., 2022b, ApJ, 925, 30
Mac Low M.-M., Klessen R. S., 2004, Rev. Mod. Phys., 76, 125
Martin-Alvarez S., Devriendt J., Slyz A., Sijacki D., Richardson M. L. A., Katz H., 2022, MNRAS, 513, 3326
Masson J., Chabrier G., Hennebelle P., Vaytet N., Commerçon B., 2016, A&A, 587, A32
McGuiness R., Smith R. J., Whitworth D. J., 2025, INPREP
Mestel L., 1966, MNRAS, 133, 265
Mocz P., Burkhart B., Hernquist L., McKee C. F., Springel V., 2017, ApJ, 838, 40
Molina F. Z., Glover S. C. O., Federrath C., Klessen R. S., 2012, MNRAS, 423, 2680
Moscadelli L., Sanna A., Beuther H., Oliva A., Kuiper R., 2022, Nature Astronomy, 6, 1068
Myers P. C., Basu S., 2021, ApJ, 917, 35
Ossenkopf V., Henning T., 1994, A&A, 291, 943
Ostriker E. C., Stone J. M., Gammie C. F., 2001, ApJ, 546, 980
Palau A., et al., 2021, ApJ, 912, 159
Passot T., Vázquez-Semadeni E., 2003, A&A, 398, 845
Pattle K., Fissel L., Tahani M., Liu T., Ntormousi E., 2023, in Inutsuka S., Aikawa Y., Muto T., Tomida K., Tamura M., eds, Astronomical Society of the Pacific Conference Series Vol. 534, Protostars and Planets VII. p. 193 (arXiv:2203.11179), doi:10.48550/arXiv.2203.11179
Pfrommer C., Werhahn M., Pakmor R., Girichidis P., Simpson C. M., 2022, MNRAS, 515, 4229
Pillai T., Kauffmann J., Tan J. C., Goldsmith P. F., Carey S. J., Menten K. M., 2015, The Astrophysical Journal, 799, 74
Planck Collaboration et al., 2016, A&A, 586, A138
Pudritz R. E., Klassen M., Kirk H., Seifried D., Banerjee R., 2014, in Petit P., Jardine M., Spruit H. C., eds, Magnetic Fields throughout Stellar Evolution Vol. 302, Magnetic Fields throughout Stellar Evolution. pp 10–20, doi:10.1017/S174392131400163X
Robinson H., Wadsley J., 2024, MNRAS, 534, 1420
Seifried D., Walch S., Weis M., Reissl S., Soler J. D., Klessen R. S., Joshi P. R., 2020, MNRAS, 497, 4196
Skalidis R., Tassis K., 2021, A&A, 647, A186
Skalidis R., Sternberg J., Beattie J. R., Pavlidou V., Tassis K., 2021, A&A, 656, A118
Soler J. D., 2019, A&A, 629, A96
Soler J. D., Hennebelle P., 2017, A&A, 607, A2
Soler J. D., et al., 2016, A&A, 596, A93
Springel V., 2010, MNRAS, 401, 791
Sur S., Federrath C., Schleicher D. R. G., Banerjee R., Klessen R. S., 2012, MNRAS, 423, 3148
Teyssier R., 2002, A&A, 385, 337
Thompson K. L., Troland T. H., Heiles C., 2019, ApJ, 884, 49
Tritsis A., Panopoulou G. V., Mouschovias T. C., Tassis K., Pavlidou V., 2015, MNRAS, 451, 4384
Troland T. H., Goss W. M., Brogan C. L., Crutcher R. M., Roberts D. A., 2016, The Astrophysical Journal, 825, 2
Vaytet N., Commerçon B., Masson J., González M., Chabrier G., 2018, A&A, 615, A5
Virtanen P., et al., 2020, Nature Methods, 17, 261
Whitworth D. J., Smith R. J., Klessen R. S., Mac Low M.-M., Glover S. C. O., Tress R., Pakmor R., Soler J. D., 2023, MNRAS, 520, 89
Whitworth D. J., et al., 2025, MNRAS, 536, 2936
Zhao B., Pudritz R. E., Pillsworth R., Robinson H., Wadsley J., 2024, ApJ, 974, 240
van der Walt S., Colbert S. C., Varoquaux G., 2011, Computing in Science Engineering, 13, 22

## APPENDIX A: BAYESIAN ANALYSIS

Bayes' Theorem states that, for a given model, the posterior distribution of the parameters $\theta = (\theta_1, \theta_2, \cdots)$ given data $D$ can be written as

$$p(\theta|D) = \frac{p(D|\theta)p(\theta)}{p(D)}, \tag{A1}$$

where $p(D|\theta)$ is the likelihood of $D$ given $\theta$, $p(\theta)$ the prior distribution of the parameters, and $p(D)$ the evidence for data $D$.

As we are analysing three distinct and different data groups, where the magnetic field strength is derived in different ways, we have to implement different variations of Bayesian analysis. We describe our methods below.

### A1 The non-hierarchical Bayesian model

*A1.1 Parameters, Likelihood and Posteriors*

As a first step, we perform the Bayesian inference with Markov Chain Monte Carlo (MCMC) sampling using the emcee code (Foreman-Mackey et al. 2013) on both the observational and numerical data in roughly the same way, with only some variation in the treatment of the errors and in the initial setup of the likelihood based on the model and the prior distributions for the parameters. In this section we present the method we implemented.

As in C10 and following works we assume a broken power-law as our underlying model, which is implemented in the likelihood module of the code. We do not consider the low density regime to be flat but allow its power-law index to vary along with all other parameters. The model

$$y_{\rm mod}(\theta) = |\mathbf{B}| \tag{A2}$$

where the magnitude of the magnetic field $|\mathbf{B}|$ is given by Eq. (2). Our parameters are thus

$$\theta = (\alpha_1, \alpha_2, n_0, B_0, \ln V), \tag{A3}$$

where $n_0$ is the number density at which the power-law changes, $\alpha_1$ and $\alpha_2$ are the power-law indexes below and above $n_0$, $B_0$ is the magnitude of the magnetic field at $n_0$, and $\ln V$ is the nuisance factor to account for intrinsic variance not described by the measurement uncertainties. The nuisance factor is only applied to the errors in $B$.

Assuming that the measurement uncertainties are Gaussian, the likelihood function given a data set $\{y_i\}$ of $n$ points

is

$$\ln p(\theta|\{y_i\}) = -\frac{1}{2}\sum_{i=1}^{n}\left[\left(\frac{y_i - y_{\mathrm{mod}}(\theta)}{s_i}\right)^2 + \ln s_i^2\right] + \mathrm{constant}, \quad (A4)$$

where $\theta$ are the free parameters, $y_{\mathrm{mod}}$ is the model as defined in Equation A2 and

$$s_i^2 = \sigma_i^2 + V^2 y_{\mathrm{mod}}^2. \quad (A5)$$

The total variance $\sigma_i$ in the measurement errors on both the magnetic field and number density is computed from

$$\sigma_i^2 = \sigma_x^2 + \sigma_y^2, \quad (A6)$$

where $\sigma_x^2$ is the error propagated from the number density $n$, computed from

$$\sigma_x^2 = \begin{cases} (\alpha_1(B_0/n_0)n^{(\alpha_1-1)}x_{err})^2 & \text{if } n \leq n_0, \\ (\alpha_2(B_0/n_0)n^{(\alpha_2-1)}x_{err})^2 & \text{if } n > n_0. \end{cases} \quad (A7)$$

and $\sigma_y^2$ is the error propagated from the magnetic field, which is computed from

$$\sigma_y^2 = y_{err}^2 + y_{model}e^{2logV}. \quad (A8)$$

This is a simple Taylor expansion for error analysis and not fully accurate, but a more detailed study is beyond the scope of this work.

For the Zeeman data we use Equation (A19) described in Appendix A3.

The posterior distribution is proportional to the product of the likelihood and the prior:

$$p(\alpha_1, \alpha_2, n_0, B_0, \ln V | y, \sigma) \propto p(y, \sigma | \alpha_1, \alpha_2, n_0, B_0, \ln V) \times$$
$$p(\alpha_1)p(\alpha_2)p(n_0)p(B_0)p(\ln V) \quad (A9)$$

where the priors are assumed to be independent.

*A1.2 A study of different priors*

C10 tested uniform priors in linear and logarithmic space for all their parameters. We first test independent uniform priors for all parameters $\theta$. We do this as we have little knowledge of how adding in a new free parameter for the diffuse gas will shape the results. Our choice for the range in $\alpha_1$ goes from -1 to 1, motivated by our lack of prior knowledge for the values of $\alpha_1$. If we were to set too narrow a limit here, we would be making assumptions about the data, possibly forcing a result within too narrow bounds. That said, we note that we are unaware of any physical mechanism that could increase the density by several orders of magnitude while simultaneously decreasing the magnetic field strength. We use the same range for $\alpha_2$ as we do not know how it will be affected by $\alpha_1$. We keep the ranges of $n_0$ and $B_0$ very broad but not unphysical and set $\ln V$ to a large range as again we have no prior knowledge on it. In summary, our priors are:

$$p(\alpha_1) = U(-1.0, 1.0)$$
$$p(\alpha_2) = U(-1.0, 1.0)$$
$$p(n_0) = U(10^{-5}, 10^5) \quad (A10)$$
$$p(B_0) = U(10^{-5}, 1000)$$
$$p(\ln V) = U(-5.0, 5.0),$$

where $U(\mathrm{min}, \mathrm{max})$ is the truncated uniform distribution.

We run the Bayesian analysis for $10^5$ steps based on 100 random walkers, remove the first $5 \times 10^4$ steps for a burn-in period, and thin to every twentieth result to avoid autocorrelation. We do this on the Zeeman line-of-sight data, taking the errors in $n$ to be a factor of 2 as in the original study (we consider different factors in Appendix A3), and present the results in Table A1. The errors are large on $\alpha_1$ and the value for $n_0$ is high. We note that the distribution of the posterior samples is not well constrained, with large scatter and unclear distributions. We therefore discard the prior given in Eq. (A10 for $\alpha_1$ for the observational data.

We next test a more constrained flat prior. Our motivation is as follows. By stating a range of -1 to 1 for both $\alpha_1$ and $\alpha_2$, we are assuming that the probability of $\alpha_1$ lying between, e.g., -1 to -0.8 is the same as of it lying between 0.8 and 1.0, which is unlikely. Taking into account previous studies that use different data, and knowledge of physically meaningful values of our parameters, we limit the priors for each parameter to:

$$p(\alpha_1) = U(-1.0, 1.0)$$
$$p(\alpha_2) = U(0.0, 1.0)$$
$$p(n_0) = U(10^{-5}, 5 \times 10^3) \quad (A11)$$
$$p(B_0) = U(10^{-5}, 100)$$
$$p(\ln V) = U(-5.0, 5.0).$$

We redo the analysis with these new priors on the Zeeman data and get results as presented in Table A1. We run the code in the same way as for the unconstrained priors and find the results to be similar. Looking at the corner plot for the Zeeman data (see Fig. A1), although the distribution is more constrained, there is a clear bimodality that can be seen in the $n_0$-$\alpha_2$ plot. It is also apparent that both $B_0$ and $n_0$ are biased towards low values, likely due to the folded normal distribution used in the likelihood function, see Appendix A3.4. What this shows is that the prior assumptions have an impact on the posterior distributions, showing that our analysis is sensitive to the choice of prior.

In order to further improve upon our assumptions and better represent the information within the data, we next tested two different Gaussian priors for $n_0$ and $B_0$ combined with a flat $p(\theta)$ for the remaining parameters. A Gaussian distribution of the form $\mathcal{N}(\mu, \sigma)$, where $\mu$ is the mean and $\sigma$ is the standard deviation, provides higher probability of the parameter value being close to $\mu$, while allowing for a gradual falloff for deviations from this value, unlike the sharp cut-off offered by the uniform distribution. For $B_0$ and $n_0$ we choose to set $\mu$ to the original values reported in C10 and $\sigma$ to be the standard deviation of the data, giving:

$$p(\alpha_1) = U(-1.0, 1.0)$$
$$p(\alpha_2) = U(-1.0, 1.0)$$
$$p(n_0) = \mathcal{N}(300, 10^6) \quad (A12)$$
$$p(B_0) = \mathcal{N}(10, 300)$$
$$p(\ln V) = U(-5.0, 5.0),$$

We also test use of the narrower distribution $\sigma = 2\mu$:

$$p(\alpha_1) = U(-0.25, 0.75)$$
$$p(\alpha_2) = U(-0.0, 1.0)$$
$$p(n_0) = \mathcal{N}(300, 600) \quad \quad \quad \quad \quad (A13)$$
$$p(B_0) = \mathcal{N}(10, 20)$$
$$p(\ln V) = U(-5.0, 5.0).$$

The results of these tests are shown in Table A1 and the right hand plot of Figure A1. With the broad Gaussian prior drawn from the dispersion of the data, $B_0$ and $n_0$ are poorly constrained, and both $\alpha_1$ and $\alpha_2$ are similar to the unconstrained uniform prior. The truncated Gaussian prior does appear to be better constrained, but care is warranted. By setting the dispersion $\sigma$ relatively small compared to the range in the data we may well be forcing the fit. The Gaussian effectively becomes a delta function returning its input value. We have tried to mitigate this by not being too restrictive, but we cannot guarantee that we have succeeded without a more detailed analysis, which is beyond the scope of this paper.

Another problem with using a Gaussian prior is that it allows for the code to predict a negative value for the variables, which is physically impossible. This is overcome however in our choice of likelihood function and error treatment discussed in Appendix A3.4, where we implement a folded normal distribution that forces the data to remain positive.

From these tests we find that the truncated uniform prior (Eq. A11), produces the most reliable fit to the Zeeman line-of-sight data. This is due to the limitations on the Gaussian prior. We use a similar uniform prior set up for the numerical simulations. In the next Appendix we describe the hierarchical Bayesian model used for the DCF data.

**A2 The hierarchical Bayesian model**

To overcome the limitations of the previous method, we employ two hierarchical Bayesian models. These models allow us to incorporate a more complete description of how the data are generated. The hierarchical models not only account for intrinsic variance in the true values of the density and magnetic field, in addition to the measurement uncertainties, but also incorporate the correlated measurement uncertainties in a straightforward manner. The increase in complexity of the models—more nuisance parameters and more complicated priors for these parameters—results in them being more computationally intensive.

For numerical stability and convergence, we define the model for the base-10 logarithms of the density and field, each scaled by the corresponding minimum data value, $\log_{10} n/n_{\min}$ and $\log_{10} B/B_{\min}$. In the following, we refer to these normalized logarithmic quantities as the density and the field.

The hierarchical models used to represent the DCF and Zeeman data are shown in Figure A2. The models assume that noisy realisations $\tilde{\mu}_X$ of the density are drawn from true values $\mu_X$ with intrinsic spread $\sigma_{X,\mathrm{int}}$. True values of the magnetic field are then generated from $\tilde{\mu}_X$ according to the power law

$$\mu_Y = \log_{10} B_0 + \alpha \, (\tilde{\mu}_X - \log_{10} n_0), \quad \quad (A14)$$

with a change in slope $\alpha$ at some break point $(n_0, B_0)$.

How the observations are drawn from the data set $(\tilde{\mu}_X, \mu_Y)$ then differs between the hierarchical model used for the DCF and the non-hierarchical model used to analyze the Zeeman observations. The magnetic field strength in the DCF data is directly correlated with number density, (as seen in Eq. 1; see Pattle et al. 2023, for a detailed discussion), and this should be accounted for in our analysis.

The measurement errors in the Zeeman model are uncorrelated, so each density and field value is drawn independently from their respective true value distributions, with a standard deviation that combines the intrinsic spread and the measurement error at that point in quadrature. The measurement errors in the DCF data, on the other hand, are correlated. Therefore, for each (density, field) pair in the DCF data, we introduce correlation between the density and field measurement errors by defining a covariance matrix

$$\Sigma = \begin{bmatrix} \sigma_{X,\mathrm{int}}^2 + \sigma_{X,\mathrm{meas}}^2 & \rho \sigma_{X,\mathrm{meas}} \sigma_{Y,\mathrm{meas}} \\ \rho \sigma_{X,\mathrm{meas}} \sigma_{Y,\mathrm{meas}} & \sigma_{Y,\mathrm{int}}^2 + \sigma_{Y,\mathrm{meas}}^2 \end{bmatrix} \quad (A15)$$

The observed density and field are then drawn from a multinormal distribution with mean $(\mu_X, \mu_Y)$ and covariance matrix $\Sigma$.

The hierarchical models were written in the `Stan` statistical modeling language[3] using version 1.2.5 of the Python package `CmdStanPy`[4]. Inference is performed using four independent Markov chains with initial values chosen to promote convergence. Each chain is run with $5 \times 10^3$ warmup iterations followed by $5 \times 10^3$ sampling iterations. The No-U-Turn Sampler algorithm is used, with parameters tuned to maintain an optimal acceptance probability (`adapt_delta = 0.99`) and to prevent excessive tree depth (`max_treedepth = 15`). We confirm convergence of the chains using the Gelman-Rubin (1992) statistic $\hat{R}$ and the bulk and tail effective sample size, all of which are provided as standard outputs by the `CmdStanPy` package's `summary` method. The posterior samples generated by the code are used to construct corner plots and compare the fits to the data, as shown in Figure 1. To evaluate the model's predictive accuracy, simulated datasets drawn from the posterior distributions can also be generated; we show the result of such predictions for the DCF data (bottom right panel in Figure 2).

**A3 Observational errors**

Before performing inference using the defined likelihood function and priors we need to address the errors on both the observational data and numerical simulations. Bayesian analysis requires error estimates on the observations as they provide the weighting for each data point. We describe the estimation and/or treatment of errors for various observational datasets in this section.

*A3.1 Errors on the DCF data*

The 700 observations compiled by Pattle et al. (2023) come from numerous different instruments including SCUBA and Planck. This implies substantial heterogeneity in their error treatments. As we do not have the raw data available to recalculate the errors consistently, we make the approximation that they were all calculated in the same manner. We do the same for the errors in $n$, assuming they are all calculated in a similar manner.

---

[3] https://mc-stan.org/
[4] https://mc-stan.org/cmdstanpy/

| Prior | $\alpha_1$ | $\alpha_2$ | $n_0$ $(10^3 \text{ cm}^{-3})$ | $B_0$ $(\mu G)$ |
|---|---|---|---|---|
| Unconstrained Uniform | $0.28^{+0.14}_{-0.17}$ | $0.72^{+0.07}_{-0.09}$ | $2.00^{+2.90}_{-1.40}$ | $2.10^{+3.90}_{-1.50}$ |
| Truncated Uniform | $0.26^{+0.09}_{-0.18}$ | $0.72^{+0.06}_{-0.06}$ | $1.80^{+1.10}_{-1.20}$ | $1.70^{+2.10}_{-1.20}$ |
| Unconstrained Gaussian | $0.54^{+0.03}_{-0.04}$ | $0.49^{+0.24}_{-0.19}$ | $850.00^{+820.00}_{-620.00}$ | $86.00^{+110.00}_{-71.00}$ |
| Truncated Gaussian | $0.16^{+0.11}_{-0.10}$ | $0.70^{+0.06}_{-0.05}$ | $0.73^{+0.48}_{-0.29}$ | $1.20^{+1.10}_{-0.80}$ |

**Table A1.** The median and 68% equal-tailed credible interval for each parameter computed from the posterior samples for the different priors tested on the Zeeman data.

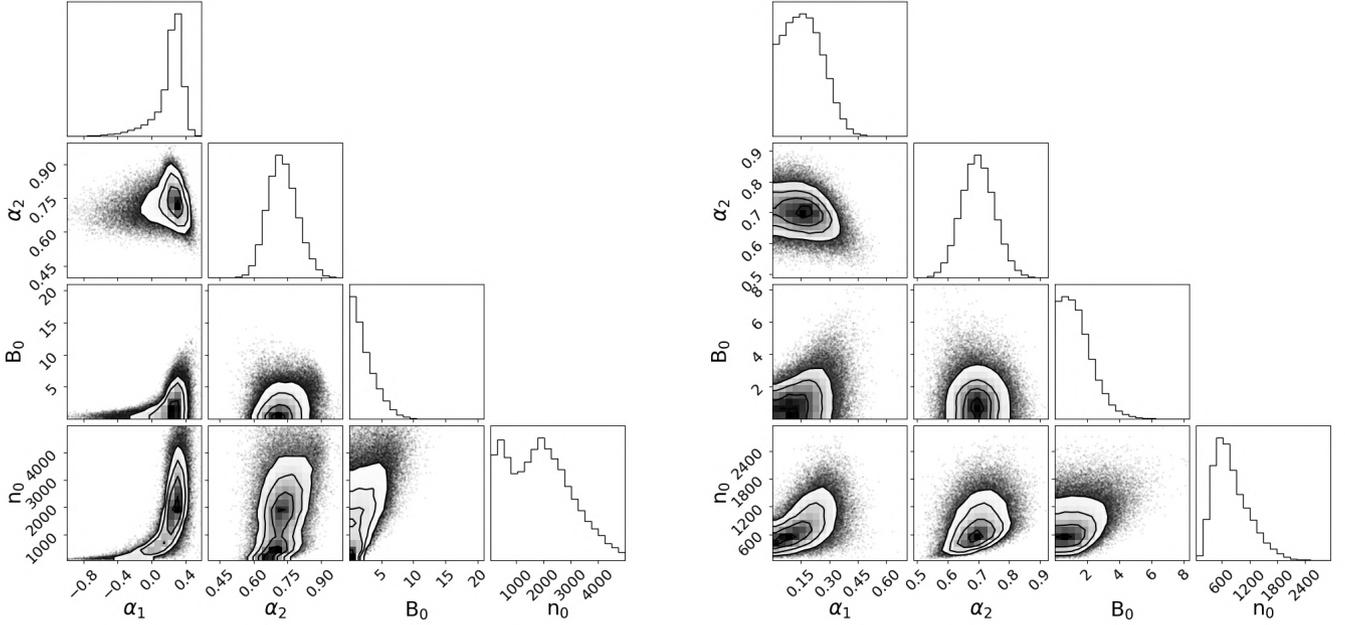

**Figure A1.** Corner plots for the Zeeman line-of-sight data using the truncated uniform prior *(left)* in Equation (A11) and the truncated Gaussian prior in Equation (A13) *(right)*. The distributions for the uniform prior are constrained but show bimodality arising in $n_0$ suggesting that there could be two break points, though these appear at densities where the Zeeman data is sparse. The truncated Gaussian prior is well constrained and appears to be a good fit, but this is due to the limits placed on the Gaussian function, which constrain the distribution too tightly.

### A3.2 DCF: Symmetrical and asymmetrical errors

Only 319 of the 700 observations have reported errors on magnetic field strength. The data without errors appears mostly at the highest densities, though they do cover the whole density range. Of the 319 observations that have errors, 26 of them have asymmetric errors and of those, five have a skew of greater than 15%. For number density there are 242 values that have reported errors out of 700. Not all of these data correspond, i.e. data that has errors in $n$ does not necessarily have reported errors in $B$ and vice versa. We select our data set for DCF based on the reported values that have symmetrical errors in $B$ and end up with a data set of 296 data points.

### A3.3 DCF: Data lacking errors

For the data that do not report errors we apply symmetric errors using the following method. We first use the data with symmetric errors to estimate the distribution of relative errors using a KDE. We then draw samples from this KDE and generate error estimates at each point that lacked errors. Thus, each data point has a reported or estimated symmetric error based on observational results. We apply this method for both magnetic field strength and number density data. Figure A3 shows the KDE for the magnetic field strength as an example.

### A3.4 Errors on the Zeeman data

The Zeeman data from C10 provides line-of-sight magnetic fields, meaning that we have both positive and negative values of the magnetic-field strength. Whilst this is not an issue for the strength, which can be taken as a magnitude, turning the negative values positive, the treatment of the errors is not as simple. In line-of-sight space the errors are symmetric and can thus be both positive and negative, but when we take the magnitude of the field strength we run into a problem. If the error is larger than the field strength then the error can return negative values of the magnitude of $|B|$, which is not physically possible. To overcome this we split the field magnitude $B$ into line-of-sight and plane of sky components such that

$$B^2 = B_z^2 + (B_x^2 + B_y)^2 \equiv B_z^2 + B_{\rm pos}^2, \tag{A16}$$

where $B_z$ and $B_{\rm pos}$ are the line-of-sight Zeeman and plane-of-sky components respectively. If only one component is measured, $B$ can be estimated by assuming that component is the average value over all possible directions (see Crutcher 2004 and Crutcher et al. 2004 for a detailed discussion of this). Note that because $B_z$ is an odd function, we average over its absolute value. Therefore,

$$B = \tfrac{4}{\pi} B_{\rm pos} \quad \text{if } B_{\rm pos} \text{ measured} \tag{A17}$$
$$B = 2|B_z| \quad \text{if } B_z \text{ measured} \tag{A18}$$

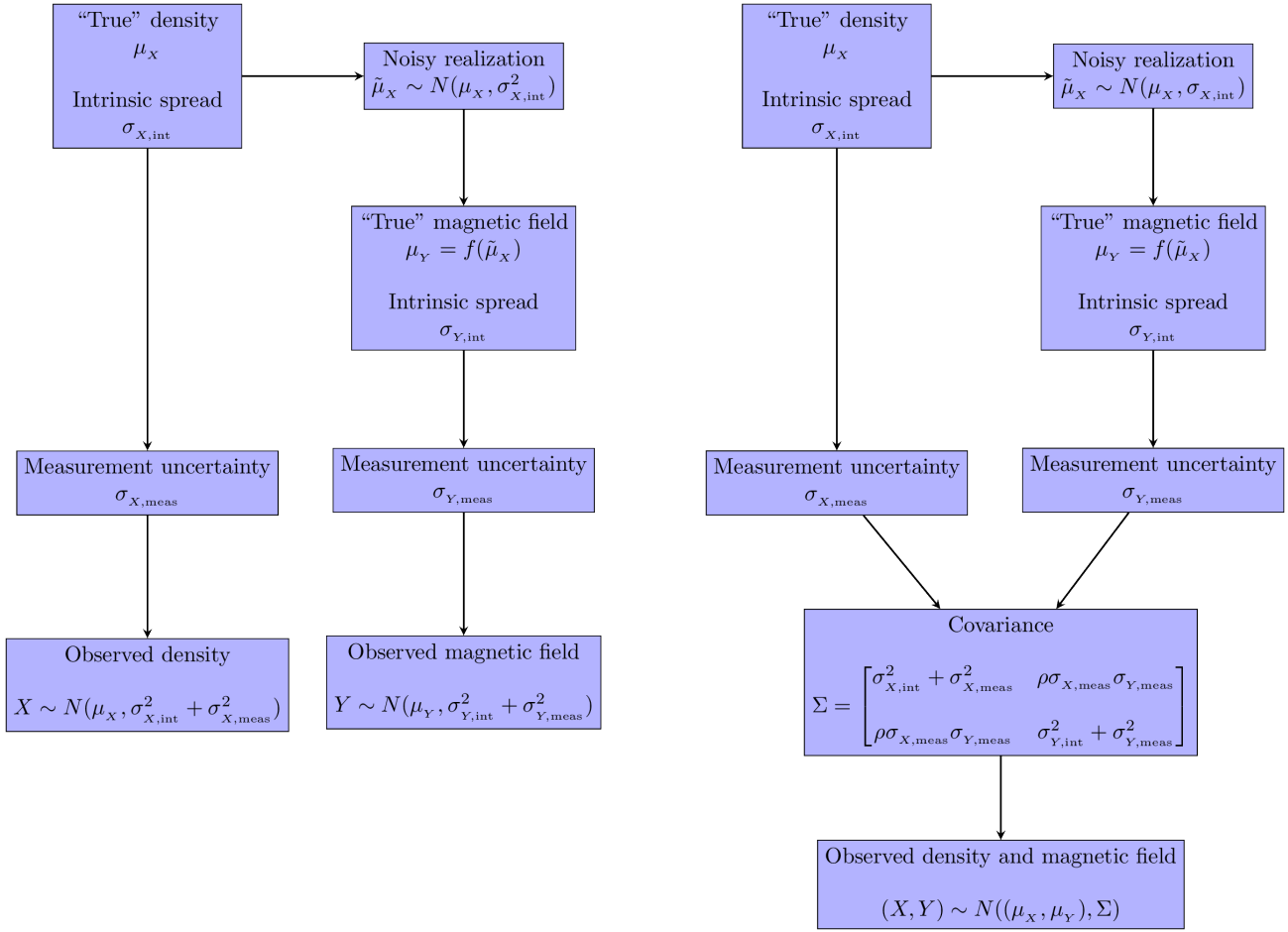

**Figure A2.** Graphical representation of the hierarchical models used for the Zeeman *(left)* and DCF *(right)* data. See text for details.

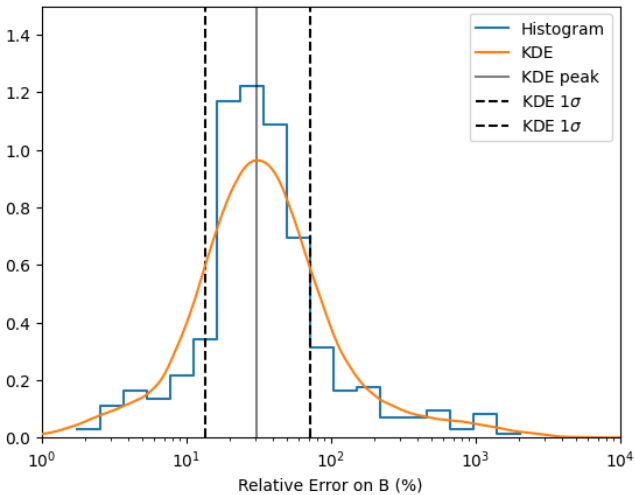

**Figure A3.** The KDE (orange line) for the distribution of relative errors, constructed using the symmetric magnetic field strength errors from the DCF data of Pattle et al. (2023). A histogram of the relative errors from the data (blue line) is also shown. The solid vertical line shows the median relative error and the dashed lines show the 1 $\sigma$ spread around it.

If only $B_{\rm pos}$ is measured, and if the measurement uncertainties are normally distributed, then $B$ is also normally distributed with mean and standard deviation $\pi/4$ times the mean and standard deviation of the $B_{\rm pos}$ measurements.[5] If only $B_z$ is measured with normally-distributed uncertainties, $B$ is then a folded normal distribution. As it is an asymmetric distribution, there is now a non-zero lower bound to the $B$ value for noisy measurements: a measurement $B_z = 0$ produces a non-zero estimate for $B$ due to the finite noise associated with the measurement. Additionally, the asymmetric distribution also implies that the error bars are no longer symmetric. We then have to compute confidence intervals from the underlying distribution.

Assuming that the factor required to convert measurements into the magnitude $B$ is $c$, the likelihood function for this problem, given a data set $\{y_i\}$ with $n$ points, is then

$$\ln p(\theta|\{y_i\}) = -\sum_{i=1}^{n}\left[\exp\left\{-\frac{1}{2}\left(\frac{y_i - y_{\rm mod}(\theta)}{cs_i}\right)^2\right\} + \right.$$
$$\left. + \exp\left\{-\frac{1}{2}\left(\frac{y_i + y_{\rm mod}(\theta)}{cs_i}\right)^2\right\} + \ln cs_i\right] + {\rm constant}, \quad ({\rm A19})$$

where $y_{\rm mod}(\theta)$ and $s_i$ are defined in Equations (A2) and (A5) respectively. As before, the latter quantity incorporates a nuisance parameter (see App. A1.1 for more details) to account for unknown systematic errors and intrinsic variance not explained by the measurement errors.

Setting $c = 2$ in Equation (A19) is equivalent to applying the conversion factor from Crutcher (2004) for Zeeman to three-dimensional field strength. If we set it to $c = 1$ we can compute the probability using the errors on the line-of-sight data. We are also able to use this principle when converting the DCF data from plane-of-sky to three dimensions, using $c = \pi/4$ instead, as described in Crutcher et al. (2004).

For the errors on $n$ we first look to the literature. C10 estimated an error of a factor of two. In follow-up work this has been brought into question as being too small and likely subject to variance across the range in density (Tritsis et al. 2015). Jiang et al. (2020) take the error as a free parameter in their study and report in their best model a factor of 9.3.

We run three tests for different error approximations using the hierarchical Bayesian method described above. We first take the original factor of two used in C10 as a Gaussian distribution where $\sigma = 2n$. We also test two other values: a factor of nine based on the work of Jiang et al. (2020), an upper limit and likely overestimation, as well as a factor of five. The results can be seen in Table A2. Whilst we do see some variation in the results, especially in $\alpha_2$, when the error factor is changed, the variations are statistically insignificant, with most lying within $1\sigma$ of each other. However, we do see that $\alpha_2$ decreases as the size of the error increases. When we go to a factor of five it is greater than $2\sigma$ from the value reported in C10 using the same error factor of two. This increases when we go to a factor of nine. The value $\alpha_2 \sim 0.50$ with an error of five shows the magnetic field could be dynamically important, whereas for a factor two, $\alpha_2 \sim 0.68$, showing the field is not dynamically important. This emphasises the importance of understanding the errors in number density and how they can change how we view the relationship and where it arises from.

---

[5] While this assumption holds for Zeeman data, according to C10, it may not hold for the plane-of-sky data if computed in a manner similar to the DCF data found in Pattle et al. (2023). We can still use the fact that $B$ will have the same distribution as $B_{\rm pos}$.

| Error Factor | $\alpha_1$ | $\alpha_2$ | $n_0$ ($\times 10^4$ cm$^{-3}$) | $B_0$ ($\mu$G) |
|---|---|---|---|---|
| **2** | | | | |
| Median | $0.14^{+0.07}_{-0.09}$ | $0.68^{+0.09}_{-0.07}$ | $0.34^{+0.55}_{-0.20}$ | $87.00^{+45.00}_{-28.00}$ |
| MAP | $0.18^{+0.04}_{-0.11}$ | $0.67^{+0.08}_{-0.08}$ | $0.18^{+0.36}_{-0.13}$ | $60.00^{+49.00}_{-12.00}$ |
| **5** | | | | |
| Median | $0.15^{+0.06}_{-0.09}$ | $0.53^{+0.09}_{-0.07}$ | $0.40^{+1.30}_{-0.30}$ | $95.00^{+60.00}_{-36.00}$ |
| MAP | $0.18^{+0.05}_{-0.10}$ | $0.50^{+0.10}_{-0.10}$ | $0.26^{+0.53}_{-0.25}$ | $61.00^{+63.00}_{-14.00}$ |
| **9** | | | | |
| Median | $0.15^{+0.06}_{-0.10}$ | $0.46^{+0.08}_{-0.06}$ | $0.33^{+1.50}_{-0.26}$ | $90.00^{+65.00}_{-33.00}$ |
| MAP | $0.18^{+0.05}_{-0.10}$ | $0.44^{+0.07}_{-0.06}$ | $0.31^{+0.44}_{-0.30}$ | $59.00^{+60.00}_{-13.00}$ |

**Table A2.** Median parameter estimates along with 68% equal-tailed intervals calculated from the posterior samples for three different relative error estimates for the number density in the Zeeman line-of-sight data. The MAP and the corresponding 68% highest posterior density (HPD) interval are also given.

Another result to consider here is the value of $n_0$. Whilst the median results stay constant across the three tests, the MAP result changes, moving to higher values. We also note that the 68% credible intervals become larger as well, meaning the determination of an exact break point becomes much less accurate. In order to fully understand the relationship, we will need more information on the uncertainties in number density.

For reporting of results we choose a factor of five, so we are not constraining them with too small a value or overestimating them with too high a value. However, we do note that, while having a full and comprehensive understanding of the errors is important for Bayesian analysis, an in-depth analysis is beyond the scope of this work.

One caveat on the method used in this work is that we are assuming all the errors are calculated in the same manner. To treat the errors properly would require each set of observations from a survey or observation to be treated independently with their own nuisance factor. The detail required for that is beyond the scope of this work.

### A4 Scaling the B field

Magnetic fields are a three-dimensional vector field and so far we have only considered the observed line-of-sight field strength for the Zeeman observations and the plane-of-sky field strengths for the DCF measurements. There are reported scaling factors for the DCF data (Crutcher 2004):

$$|\mathbf{B}| = \frac{4}{\pi} B_{\rm pos}, \quad ({\rm A20})$$

and for the Zeeman data (Crutcher et al. 2004),

$$|\mathbf{B}| = 2 B_z. \quad ({\rm A21})$$

however, if we apply these to the data we also need to apply the scaling factor to the errors. This results in just a shift in the value of $B_0$ so we do not investigate this.

Pattle et al. (2023) show that the average DCF measurement is a factor $d = 6.3 \pm 1.5$ higher than the average Zeeman measurement. This is a different conversion factor to $f$ already seen in Equation (1), which is applied to the raw data, as $d$ is specifically the ratio between the DCF and Zeeman data. We perform a fit with this discrepancy taken into account where we reduce only the DCF measurements by this factor $d$. We do not scale the errors, as they are likely understated, since the DCF over-predicts the field strength. Not reducing the errors also allows for any physical variability in the scale factor.

| Prior | $\alpha_1$ | $\alpha_2$ | $n_0$ ($\times 10^4$ cm$^{-3}$) | $B_0$ ($\mu$G) |
|---|---|---|---|---|
| **DCF scaled** | | | | |
| Median | $0.26^{+0.03}_{-0.03}$ | $0.63^{+0.12}_{-0.10}$ | $1.81^{+3.58}_{-1.29}$ | $281.00^{+154.00}_{-100.00}$ |
| MAP | $0.23^{+0.61}_{-0.01}$ | $0.58^{+0.15}_{-0.06}$ | $0.81^{+0.81}_{-0.81}$ | $1.90^{+1.87}_{-1.87}$ |

**Table A3.** The median and 68% credible interval and the MAP with its corresponding 68% HPD interval for each parameter from the posterior samples for the DCF data, where the magnetic field strength has been scaled down using the factor of $d = 6.3$ from (Pattle et al. 2023) to match Zeeman magnetic field strengths.

| Data set | $\alpha_1$ | $\alpha_2$ | $n_0$ ($\times 10^4$ cm$^{-3}$) | $B_0$ ($\mu$G) |
|---|---|---|---|---|
| **DCF** | | | | |
| Median | $0.26^{+0.01}_{-0.02}$ | $0.77^{+0.14}_{-0.15}$ | $13.90^{+10.10}_{-7.40}$ | $463.00^{+117.00}_{-104.00}$ |
| MAP | $0.26^{+0.01}_{-0.02}$ | $0.78^{+0.15}_{-0.14}$ | $12.00^{+8.00}_{-8.00}$ | $440.00^{+130.00}_{-93.00}$ |
| **DCF + outlier** | | | | |
| Median | $0.26^{+0.01}_{-0.02}$ | $0.64^{+0.12}_{-0.10}$ | $8.80^{+8.70}_{-4.60}$ | $413.00^{+121.00}_{-93.00}$ |
| MAP | $0.26^{+0.01}_{-0.02}$ | $0.62^{+0.11}_{-0.10}$ | $5.50^{+7.70}_{-3.20}$ | $370.00^{+140.00}_{-62.00}$ |
| **DCF$_{\text{kde}}$** | | | | |
| Median | $0.35^{+0.11}_{-0.11}$ | $0.04^{+0.05}_{-0.03}$ | $470.00^{+350.00}_{-200.00}$ | $3600.00^{+620.00}_{-590.00}$ |
| MAP | $0.35^{+0.01}_{-0.01}$ | $0.01^{+0.01}_{-0.01}$ | $360.00^{+300.00}_{-180.00}$ | $3600.00^{+570.00}_{-630.00}$ |

**Table A4.** Median parameter estimates plus their 68% credible intervals and the MAPs for the observational *B-n* relation for the DCF data set with and without the high density outlier and for the full DCF data that includes the 404 data that are missing errors. The missing errors are calculated using a KDE. The value of $n_0$ shifts higher, while the other parameters are relatively unchanged.

We run the hierarchical Bayesian analysis on the scaled-down DCF data, with results given in Table A3. The scaling factor applied here is just a ratio between the two data sets and therefore, not well constrained and possibly not representative of the true scaling. Therefore we present this for interest only and note that due to inherent biases these results are also likely to be skewed. We see no statistical variation in the parameters, except in $B_0$, which is to be expected as this is the only value to have changed in the data.

### A5 Influential data points

Figure A4 shows two versions of the DCF data set. *Right* the full data set with an outlier at $n > 10^8$ cm$^{-3}$. As previously discussed, the DCF method becomes inaccurate at extremely high number densities. *Left* the same data set, excluding the high density outlier. Table A4 shows that when we exclude the outlier the inferred position of the break changes to a higher density. Due to the uncertain nature of this point we exclude it from our main result.

Our analysis is limited by the fact that the errors in the underlying data sets are mostly broad estimates, with some data completely lacking errors. The lack of errors particularly affects the densest gas ($n > 10^7$ cm$^{-3}$) where Pattle et al. (2023) suggested there could be a turn over and flattening of the relationship. It is also at these densities where the DCF technique becomes difficult and less reliable. This is discussed in more detail in Section 2.1.

The results of an analysis on the full data set are shown in Figure A5, where we have used a KDE to estimate the errors on the 404 DCF data points missing errors (see App. A3 for a detailed description). This figure shows a turn over and reduction at high densities as predicted in Pattle et al. (2023). They raise the possibility of an additional transition point. However, that point would occur at densities where the DCF becomes unreliable due to radiation pressure (Lazarian 2020). Unfortunately, these high density points do not have reported errors and are likely to be biasing the result. We therefore do not consider them in our main results.

### A6 Inference on the numerical data

An issue in running the Bayesian routine on the numerical simulations is the large number of data points, with each simulation consisting of well over five million points. Running the routine on such large data sets takes substantial resources. To mitigate the computational cost, we take five random samples of $10^5$ points each from each simulation, using the likelihood function Equation (A4) with priors

$$\begin{aligned}
p(\alpha_1) &= U(-1.0, 1.0) \\
p(\alpha_2) &= U(-1.0, 1.0) \\
p(n_0) &= U(0.0, 10^6) \\
p(B_0) &= U(10^{-5}, 1000) \\
p(\ln V) &= U(-5.0, 5.0),
\end{aligned} \quad \text{(A22)}$$

The initial conditions for the Bayesian analyses are taken from a linear regression fit of the whole data set from a simulation and used for each random sample,

$$(\alpha_1, \alpha_2, B_0, n_0, \ln V) = (0.34, 0.34, 10.0, 300.0, 0.5) \quad \text{(A23)}$$

We take 100 walkers and run $2.50 \times 10^4$ steps to ensure convergence. We then combine the five individual data sets for each simulation into one data set, discarding the first $5 \times 10^3$ steps and thin to every twentieth data point.

Once the code has been executed, we have five sets of posterior parameter samples for each simulation. We combine these five sets of samples to compute the median and MAP values as well as the corresponding 68% credible intervals for each parameter and present them in Figure 3 and in Appendix A8.

All simulations bar the SSD and LSD simulations of Gent et al. (2024) use the same initial conditions as the non-hierarchical Bayesian technique used on the observational data seen in Section A1.1. The reason for the SSD simulations differing is that the data in these diffuse ISM simulations all lie below the values used for the initial condition and so the code does not converge as there is no data to constrain it.

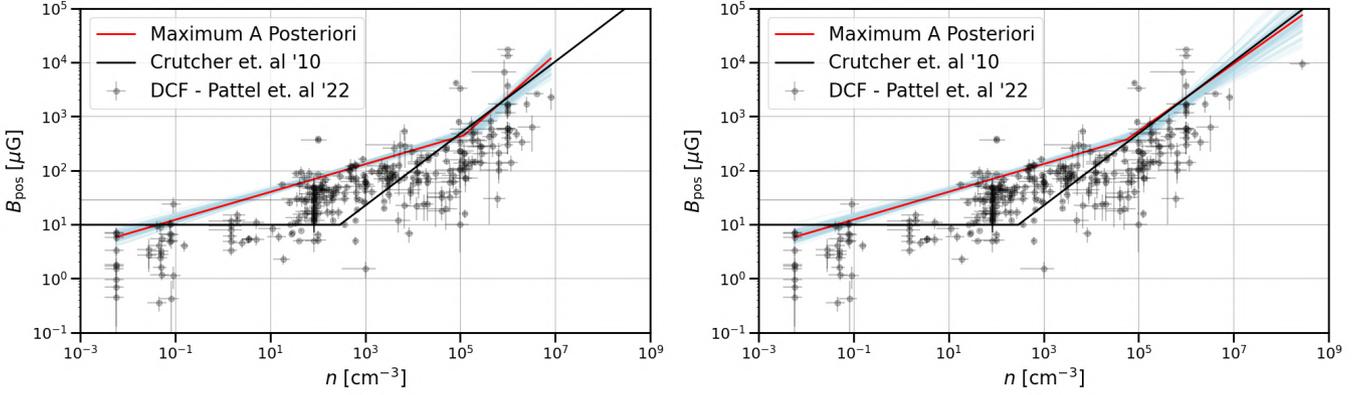

**Figure A4.** Results from observations of magnetic field strength as a function of number density. Left: The DCF data set from Pattle et al. (2023) with the high density outlier removed. Right: The DCF data set from Pattle et al. (2023) including the high density outlier. The *blue lines* on these plots are models computed for 100 posterior parameter samples from our analysis. The *solid black line* on all plots is the relationship proposed by C10.

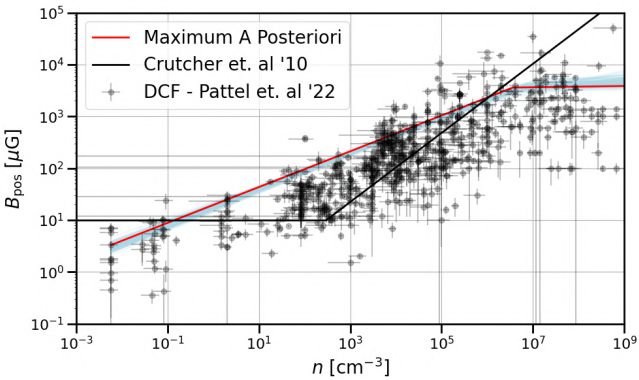

**Figure A5.** Plot showing the results of the hierarchical Bayesian analysis on the full DCF data set from Pattle et al. (2023) where missing errors have been calculated using a KDE. We can see here that the relationship does not match the original and has a turn over at the highest densities. These are the densities where it is expected that the DCF method becomes inaccurate due to radiation pressure. We also find that most of the high density points were not reported with errors, introducing uncertainty to the quality of this fit.

For these four simulations, we instead use:

$$(\alpha_1, \alpha_2, n_0, B_0, \ln V) = (0.0, 0.0, 0.1, 1.0, 0.5). \tag{A24}$$

### A7 Errors on the numerical data

The numerical data from the various simulations do not include error estimates. Whilst there are various methods of applying errors to the data, we choose a simple extension from that used in the observational data. We take the KDE errors calculated from the observational data for both the magnetic field strength and number density in Appendix A3.3 and apply them to the numerical simulations. As with the observational data, we combine a nuisance factor with these measurement errors.

### A8 Tabulated numerical results tables

We provide the results from the numerical simulation data plotted in Figure 3 in tabular form in Tables A5 and A6.

## APPENDIX B: HIGH RESOLUTION MHD SIMULATION OF DWARF GALAXIES

To measure the importance of numerical resolution effects, we re-simulated MHD_10 and MHD_SAT from W23 with higher resolution parameters given in Table B1. The new simulations attain resolutions of $r_{\rm cell} \approx 0.01$ pc at $n \approx 10^6$ cm$^{-3}$ (see Fig. B1). In the diffuse gas, $n \approx 10^{-2}$–1 cm$^{-3}$ and the resolution is below $r_{\rm cell} < 10$ pc, similar to the resolution of the densest gas in the RAMSES Milky Way-like simulations. The new base gas mass in the cells is set to 1 M$_\odot$, allowed to vary by a factor of two, and further refined to resolve the Jeans length, which is set to eight cells. The accretion radius around the sink particles is reduced to 0.01 pc. The maximum mass that a sink is allowed to reach is set to 1000 M$_\odot$, allowing for massive star formation. We have also included the effects of photoionization. The method implemented here is described in Whitworth et al. (2025). The new simulations are run for 10 Myr from the 1 Gyr checkpoint of the original simulations to allow for the new resolution to take effect, a new population of sink particles to have formed, and ensure that no large-scale changes in the behaviour or growth of the magnetic field have arisen (see Fig. B2).

## APPENDIX C: THE DIFFUSE GAS PHASE

We briefly examine whether the behaviour of the field changes when we better resolve the diffuse gas. We compare the plasma $\beta = P/P_{\rm mag}$, the ratio of the thermal to the magnetic pressure, in the SMC like dwarf galaxy simulations of Whitworth et al. (2023) and the high resolution models discussed in Appendix B. When $\beta < 1$, magnetic pressure dominates over thermal pressure. If resolution does affect how the field behaves in the diffuse gas we would expect to see lower $\beta$ in the diffuse gas for the high resolution model.

Figure C1 shows the difference between the simulations in plasma-$\beta$. At high densities magnetic pressure dominates over thermal pressure. At lower densities it is less clear. To be able to see the change we compared the mass fraction of gas with $\beta < 1$ between the simulations. There is no change in the gas mass fraction with $n < 300$ cm$^{-3}$ and $\beta < 1$. However when we look at gas with $n \leq 1$ cm$^{-3}$ the gas mass increases by a factor of 2.9 for MHD_10 and a factor of 0.38 for MHD_SAT. So we are seeing more diffuse gas being magnetically supported, though not a significant increase. This is likely due to the fact that the diffuse gas is still not resolved enough for significant changes to be seen. The high resolution models have cell radii of a few parsecs at $n \sim 1$ cm$^{-3}$ which is too large to capture fast growth of the SSD, whilst the original simulations have cell radii of $\sim 10$ pc at $n \sim 1$ cm$^{-3}$.

| Simulation | $\alpha_1$ | $\alpha_2$ | $n_0$ (cm$^{-3}$) | $B_0$ ($\mu$G) | Section |
|---|---|---|---|---|---|
| NI | $-0.21^{+0.02}_{-0.02}$ | $0.56^{+0.01}_{-0.00}$ | $690.00^{+11.00}_{-10.00}$ | $15.00^{+0.00}_{-0.00}$ | 4.1 |
| IMHD | $-0.27^{+0.01}_{-0.01}$ | $0.56^{+0.00}_{-0.00}$ | $710.00^{+21.00}_{-13.00}$ | $14.00^{+0.00}_{-0.00}$ | 4.1 |
| B3 | $0.57^{+0.00}_{-0.00}$ | $0.20^{+0.00}_{-0.00}$ | $0.45^{+0.01}_{-0.02}$ | $1.80^{+0.00}_{-0.00}$ | 4.2 |
| B6 | $0.67^{+0.00}_{-0.01}$ | $0.15^{+0.00}_{-0.00}$ | $0.26^{+0.01}_{-0.01}$ | $2.10^{+0.00}_{-0.00}$ | 4.2 |
| KPC_LOW | $0.13^{+0.01}_{-0.01}$ | $0.36^{+0.01}_{-0.01}$ | $1.10^{+0.10}_{-0.10}$ | $3.10^{+0.10}_{-0.00}$ | 4.2 |
| KPC_HIGH | $0.23^{+0.00}_{-0.00}$ | $0.13^{+0.01}_{-0.02}$ | $21.00^{+4.00}_{-3.00}$ | $11.00^{+0.00}_{-0.00}$ | 4.2 |
| SSD_35 | $0.30^{+0.0}_{-0.10}$ | $-1.00^{+0.00}_{-0.00}$ | $0.16^{+0.23}_{-0.04}$ | $0.22^{+0.03}_{-0.02}$ | 4.2 |
| SSD_98 | $0.67^{+0.02}_{-0.01}$ | $-0.15^{+0.70}_{-0.01}$ | $0.058^{+0.00}_{-0.06}$ | $0.29^{+0.01}_{-0.28}$ | 4.2 |
| SSD_700 | $0.69^{+0.00}_{-0.01}$ | $0.39^{+0.01}_{-0.01}$ | $0.019^{+0.00}_{-0.00}$ | $0.50^{+0.00}_{-0.00}$ | 4.2 |
| LSD_3500 | $0.78^{+0.00}_{-0.01}$ | $0.01^{+0.00}_{-0.00}$ | $0.16^{+0.01}_{-0.00}$ | $10.00^{+0.00}_{-0v}$ | 4.2 |
| MHD_10 | $0.49^{+0.00}_{-0.00}$ | $0.44^{+0.02}_{-0.02}$ | $360.00^{+71.00}_{-110.00}$ | $11.00^{+1.00}_{-2.00}$ | 4.3.1 |
| MHD_SAT | $0.33^{+0.00}_{-0.00}$ | $0.39^{+0.00}_{-0.00}$ | $10.00^{+5.00}_{-3.00}$ | $3.40^{+0.50}_{-0.40}$ | 4.3.1 |
| MHD_10_HR | $0.48^{+0.00}_{-0.01}$ | $0.49^{+0.00}_{-0.01}$ | $700.00^{+1600.00}_{-340.00}$ | $19.00^{+14.00}_{-5.00}$ | 4.3.1 |
| MHD_SAT_HR | $0.28^{+0.00}_{-0.00}$ | $0.54^{+0.01}_{-0.03}$ | $210.00^{+25.00}_{-92.00}$ | $7.20^{+0.20}_{-0.90}$ | 4.3.1 |
| MHD_10_LR | $0.84^{+0.03}_{-0.22}$ | $0.52^{+0.01}_{-0.19}$ | $0.63^{+2900.00}_{-0.07}$ | $0.09^{+13.00}_{-0.00}$ | 4.3.1 |
| MW_RAM_1 | $0.56^{+0.00}_{-0.00}$ | $0.14^{+0.04}_{-0.03}$ | $22.00^{+2.00}_{-2.00}$ | $48.00^{+2.00}_{-2.00}$ | 4.3.2 |
| MW_RAM_2 | $0.72^{+0.02}_{-0.02}$ | $0.46^{+0.00}_{-0.00}$ | $0.40^{+0.0}_{-0.10}$ | $1.30^{+0.00}_{-0.10}$ | 4.3.2 |
| MWR_R1 | $0.59^{+0.10}_{-0.33}$ | $0.52^{+0.01}_{-0.01}$ | $17.00^{+6.00}_{-14.00}$ | $12.00^{+2.00}_{-8.00}$ | 4.3.2 |
| MWR_R2 | $0.61^{+0.01}_{-0.01}$ | $0.44^{+0.00}_{-0.00}$ | $38.00^{+2.00}_{-1.00}$ | $16.00^{+0.00}_{-0.00}$ | 4.3.2 |

**Table A5.** The median posterior parameter values in the $B$-$n$ relation inferred from five sets of random samples from the numerical simulation data, along with the corresponding 68% credible intervals. The last column in every row links to the section where we discuss the corresponding results. In some cases, large relative errors in the inferred parameters arise from outliers due to our random sampling of the data sets. We note the small errors on the exponents $\alpha_1$ and $\alpha_2$ showing that they are well constrained for each simulation. It is clear from these results that the relationship changes dramatically across simulations, with large variations in all free parameters.

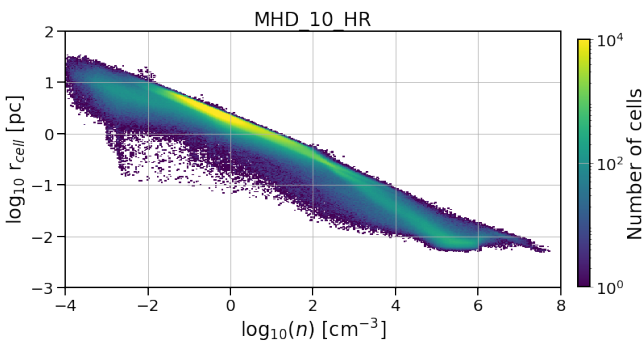

**Figure B1.** The cell resolution of simulation MHD_10_HR where we can see sub-parsec resolution being reached around number densities of $n \sim 1$ cm$^{-3}$.

Numerical simulations tend not to resolve the diffuse medium for computational efficiency. Galaxy evolution simulations also do not consider the diffuse medium as important as no star formation takes place there. However, the magnetic field in the diffuse medium is dynamically important as it will slow down accretion onto giant molecular clouds, slowing their growth. Sufficiently high numerical resolution is needed to properly simulate these effects (see Martin-Alvarez et al. 2022, for further resolution studies).

| Simulation | $\alpha_1$ | $\alpha_2$ | $n_0$ (cm$^{-3}$) | $B_0$ ($\mu$G) | Section |
|---|---|---|---|---|---|
| NI | $-0.20^{+0.00}_{-0.00}$ | $0.56^{+0.00}_{-0.00}$ | $690.00^{+10.00}_{-10.00}$ | $14.00^{+1.00}_{-0.00}$ | 4.1 |
| IMHD | $-0.27^{+0.01}_{-0.01}$ | $0.55^{+0.00}_{-0.00}$ | $700.00^{+21.00}_{-10.00}$ | $14.00^{+0.00}_{-0.00}$ | 4.1 |
| B3 | $0.57^{+0.01}_{-0.00}$ | $0.20^{+0.00}_{-0.00}$ | $0.45^{+0.01}_{-0.02}$ | $1.80^{+0.00}_{-0.00}$ | 4.2 |
| B6 | $0.68^{+0.00}_{-0.01}$ | $0.15^{+0.00}_{-0.00}$ | $0.23^{+0.04}_{-0.03}$ | $2.10^{+0.00}_{-0.00}$ | 4.2 |
| KPC_LOW | $0.13^{+0.01}_{-0.01}$ | $0.36^{+0.01}_{-0.01}$ | $1.10^{+0.0}_{-0.10}$ | $3.10^{+0.00}_{-0.00}$ | 4.2 |
| KPC_HIGH | $0.23^{+0.00}_{-0.00}$ | $0.13^{+0.02}_{-0.01}$ | $20.00^{+4.00}_{-3.00}$ | $1.00^{+0.00}_{-0.00}$ | 4.2 |
| SSD_35 | $0.30^{+00.0}_{-0.00}$ | $-1.00^{+0.00}_{-0.00}$ | $0.16^{+0.00}_{-0.04}$ | $0.22^{+0.00}_{-0.02}$ | 4.2 |
| SSD_98 | $0.67^{+0.01}_{-0.01}$ | $-0.16^{+0.32}_{-0.03}$ | $0.06^{+0.00}_{-0.02}$ | $0.29^{+0.01}_{-0.07}$ | 4.2 |
| SSD_700 | $0.69^{+0.01}_{-0.00}$ | $0.39^{+0.00}_{-0.01}$ | $0.02^{+0.01}_{-0.00}$ | $0.50^{+0.00}_{-0.00}$ | 4.2 |
| LSD_3500 | $0.78^{+0.00}_{-0.00}$ | $0.01^{+0.00}_{-0.00}$ | $0.17^{+0.00}_{-0.00}$ | $10.00^{+0.00}_{-0.00}$ | 4.2 |
| MHD_10 | $0.49^{+0.00}_{-0.00}$ | $0.45^{+0.02}_{-0.02}$ | $390.00^{+66.00}_{-100.00}$ | $12.00^{+1.00}_{-2.00}$ | 4.3.1 |
| MHD_SAT | $0.34^{+0.00}_{-0.01}$ | $0.39^{+0.00}_{-0.00}$ | $8.10^{+5.10}_{-2.10}$ | $3.10^{+0.70}_{-0.20}$ | 4.3.1 |
| MHD_10_HR | $0.48^{+0.00}_{-0.01}$ | $0.49^{+0.01}_{-0.00}$ | $550.00^{+580.00}_{-360.00}$ | $16.00^{+8.00}_{-5.00}$ | 4.3.1 |
| MHD_SAT_HR | $0.28^{+0.00}_{-0.00}$ | $0.55^{+0.01}_{-0.01}$ | $220.00^{+26.00}_{-35.00}$ | $7.40^{+0.30}_{-0.40}$ | 4.3.1 |
| MHD_10_LR | $0.85^{+0.04}_{-0.05}$ | $0.52^{+0.02}_{-0.02}$ | $2.00^{+1.00}_{-2.00}$ | $0.10^{+0.00}_{-0.02}$ | 4.3.1 |
| MW_RAM_1 | $0.56^{+0.00}_{-0.00}$ | $0.14^{+0.04}_{-0.03}$ | $22.00^{+2.00}_{-2.00}$ | $48.00^{+2.00}_{-2.00}$ | 4.3.2 |
| MW_RAM_2 | $0.72^{+0.03}_{-0.00}$ | $0.46^{+0.00}_{-0.00}$ | $0.43^{+0.01}_{-0.05}$ | $1.30^{+0.10}_{-0.00}$ | 4.3.2 |
| MWR_R1 | $0.59^{+0.13}_{-0.02}$ | $0.52^{+0.00}_{-0.01}$ | $15.00^{+8.00}_{-1.00}$ | $12.00^{+2.00}_{-1.00}$ | 4.3.2 |
| MWR_R2 | $0.62^{+0.01}_{-0.01}$ | $0.44^{+0.00}_{-0.00}$ | $38.00^{+2.00}_{-1.00}$ | $16.00^{+0.00}_{-0.00}$ | 4.3.2 |

**Table A6.** The MAP values in the $B$-$n$ relation inferred from five sets of random samples from the numerical simulation data, along with the corresponding 68% HPD. The section where we discuss these results is also specified. In some cases, large relative errors in the inferred parameters arise from outliers due to our random sampling of the data sets. We note the small errors on the exponents $\alpha_1$ and $\alpha_2$ showing that they are well constrained for each simulation. It is clear from these results that the relationship changes dramatically across simulations, with large variations in all free parameters.

| | |
|---|---|
| $\rho_c$ (g cm$^{-3}$) | $2.40 \times 10^{-19}$ |
| $n$ (cm$^{-3}$) | $10^5$ |
| $r_{acc}$ (pc) | 0.01 |
| Softening Length (pc) | 0.10 |
| Max sink mass (M$_\odot$) | 1000 |
| Base cell mass (M$_\odot$) | 1 |

**Table B1.** Sink particle and resolution parameters used for the new AREPO MHD simulations, with $\rho_c$ the sink density threshold, $n$ the number density of the sink threshold, and $r_{acc}$ the accretion radius. The softening length is the adaptive gravitational softening length of the gas cells and the base cell mass is the target mass of the cell, which is allowed to vary by factor of two.

This paper has been typeset from a T$_E$X/L$^A$T$_E$X file prepared by the author.

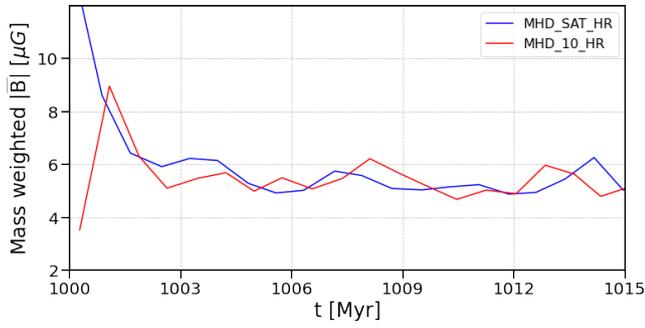

**Figure B2.** The mass-weighted absolute magnetic field strength in simulations MHD_10_HR and MHD_SAT_HR taken from within a radius of $r = 1.5$ kpc and height $z = \pm 0.2$ pc. We see no change in the field strength over the 15 Myr of the new simulations. The peak at the beginning is from the code refining the gas to the new levels.

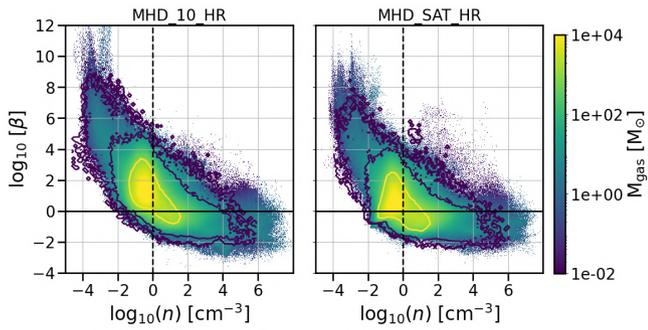

**Figure C1.** Plasma $\beta$ versus number density for MHD_10_HR and MHD_SAT_HR. The contours show the same but for simulations MHD_10 and MHD_SAT. Their colours correspond to the values on the colour bar. The *black horizontal line* shows $\beta = 1$. Below this magnetic pressure dominates over thermal pressure. In the denser regions to the right of the *black vertical dashed line* at $n = 1$ cm$^{-3}$, we see more gas that is magnetically dominated. In the low density gas there is no clear difference.